\documentclass[preprint,onecolumn,nofootinbib]{revtex4}
\usepackage[colorlinks=true,linkcolor=blue,urlcolor=blue,filecolor=black,citecolor=red,pdfstartview=FitV,pdftitle={},pdfsubject={},pdfkeywords={},pdfpagemode=None,bookmarksopen=true]{hyperref}
\usepackage{graphicx}
\usepackage{amsmath}
\usepackage{amsfonts}
\usepackage{amssymb,ulem}
\usepackage{color,xcolor}%
\usepackage{CJK}
\usepackage{subfigure}
\usepackage{amsthm,amsmath,amssymb}
\usepackage{mathrsfs}
\usepackage{multirow}

\usepackage{dcolumn}
\usepackage{float}
\usepackage{amsmath}
\usepackage{mathrsfs} 

\newcommand{\red}{\textcolor[rgb]{1.00,0.00,0.00}}

\begin{document}
\title{Peculiar properties in quasi-normal spectra from loop quantum gravity effect}
\author{Guoyang Fu$^{1,2,5}$}
\thanks{FuguoyangEDU@163.com}
\author{Dan Zhang $^{3}$}
\thanks{danzhanglnk@163.com}
\author{Peng Liu $^{4}$}
\thanks{phylp@email.jnu.edu.cn}
\author{Xiao-Mei Kuang$^{2}$}
\thanks{xmeikuang@yzu.edu.cn}
\author{Jian-Pin Wu$^{2}$}
\thanks{jianpinwu@yzu.edu.cn, corresponding author}
\affiliation{$^1$ Department of Physics and Astronomy, Shanghai Jiao Tong University, Shanghai 200240, China}
\affiliation{$^2$ Center for Gravitation and Cosmology, College of Physical Science and Technology, Yangzhou University, Yangzhou 225009, China}
\affiliation{$^3$~ Key Laboratory of Low Dimensional Quantum Structures and Quantum Control of Ministry of Education, Synergetic Innovation Center for Quantum Effects and Applications, and Department of Physics, Hunan Normal University, Changsha, Hunan 410081, China}
\affiliation{$^4$ Department of Physics and Siyuan Laboratory, Jinan University, Guangzhou 510632, P.R. China}
\affiliation{$^5$ Shanghai Frontier Science Center for Gravitational Wave Detection, Shanghai Jiao Tong University, Shanghai 200240, China}

\begin{abstract}
\baselineskip=0.4 cm

We investigate the quasi-normal mode (QNM) spectra for scalar and electromagnetic fields over a covairant loop quantum gravity black hole (LQG-BH). For the fundamental modes, the LQG effect reduces the oscillations in the scalar field, however it induces stronger oscillations in the electromagnetic field, comparing to the classical case. Under the scalar field perturbation, the system enjoys faster decaying modes with more oscillations than the electromagnetic field. Some peculiar phenomena emerge in the QNM spectra with higher overtones. A notable feature is the substantial divergence observed in the first several overtones from their Schwarzschild counterparts, with this discrepancy magnifying as the overtone number increases. Another remarkable phenomenon in higher overtones is that the QNFs of the scalar field with $l=0$ exhibit an oscillatory behavior as the quantum parameter $r_0$ increases significantly. These oscillations intensify with the rising overtone number.  We hypothesize that this oscillatory pattern may be associated with the extremal effect.

\end{abstract}

\maketitle

\section{Introduction}\label{sec-intro}

A non-perturbative and background-independent technique, loop quantum gravity (LQG) \cite{Rov,Thiemann:2001gmi,Ashtekar:2004eh,Han:2005km}, provides a scenario for quantizing space-time structure. This approach has been successfully applied to quantize symmetry reduced cosmological space-times, known as loop quantum cosmology (LQC) \cite{Bojowald:2001xe,Ashtekar:2006rx,Ashtekar:2006uz,Ashtekar:2006wn,Ashtekar:2003hd,Bojowald:2005epg,Ashtekar:2011ni,Wilson-Ewing:2016yan}. Effective LQC theory can be constructed by incorporating two key quantum gravity effects, namely the inverse volume correction and the holonomy correction, which can be achieved using both the canonical approach \cite{Taveras:2008ke,Ding:2008tq,Yang:2009fp,Bojowald:2009jj,Bojowald:2009jk,Bojowald:2010qm} and the path integral perspectives \cite{Ashtekar:2009dn,Ashtekar:2010ve,Ashtekar:2010gz,Huang:2011es,Qin:2012gaa,Qin:2011hx,Qin:2012xh}.
The quantum gravity effects in LQC can be connected to low-energy physics, resulting in a solvable cosmological model for studying quantum gravity effects. In particular, the big bang singularity in classical general relativity (GR) is successfully avoided by the quantum gravity effects \cite{Bojowald:2001xe,Ashtekar:2006rx,Ashtekar:2006uz,Ashtekar:2006wn,Ashtekar:2003hd,Bojowald:2005epg,Ashtekar:2011ni,Wilson-Ewing:2016yan,Bojowald:2003xf,Singh:2003au,Vereshchagin:2004uc,Date:2005nn,Date:2004fj,Goswami:2005fu}, which instead result in a non-singular big bounce even at the semi-classical level \cite{Bojowald:2005zk,Stachowiak:2006uh}.

Following the same idea in LQC \cite{Bojowald:2001xe,Ashtekar:2006rx,Ashtekar:2006uz,Ashtekar:2006wn,Ashtekar:2003hd,Bojowald:2005epg,Ashtekar:2011ni,Wilson-Ewing:2016yan}, several effective black holes (BH) models with LQG corrections have been constructed. Up to date, most of effective LQG-BHs are implemented through the input of the holonomy correction; see, for example, \cite{Ashtekar:2005qt,Modesto:2005zm,Modesto:2008im,Campiglia:2007pr,Bojowald:2016itl,Boehmer:2007ket,Chiou:2008nm,Chiou:2008eg,Joe:2014tca,Yang:2022btw,Gan:2022oiy} and references therein. A common feature of LQG-BHs is that the singularity is replaced by a transition surface between a trapped and an anti-trapped region, which can be understood as the interior region of black hole and white hole.

The heart of the holonomy correction is the phase space regularisation technique called polymerisation \cite{Corichi:2007tf}. Because of this, the effective LQG-BH with holonomy correction is also known as the polymer BH. The basic idea behind polymerisation is the replacement of the conjugate momentum $p$ with their regularised counterpart $\sin(\bar{\lambda} p)/\bar{\lambda}$, where $\bar{\lambda}$ is a quantity known as polymerisation scale, which is linked to the area-gap. Depending on whether the polymerization scale is constant or phase space dependent function, the polymer BHs are classified into two basic types:
\begin{itemize}
\item \textit{$\mu_0$-type scheme}

In this scheme, the polymerization scale is assumed to remain constant over the whole phase space \cite{Ashtekar:2005qt,Modesto:2005zm,Modesto:2008im,Campiglia:2007pr,Bojowald:2016itl}. This approach has the drawback that the final result is reliant on the fiducial structures, which are introduced in the construction of the classical phase space. In addition, even in the low-curvature regimes, significant quantum effects may manifest, making these models unphysical. To overcome this drawback, some generalized version of the $\mu_0$-scheme have been proposed, see for example \cite{Corichi:2015xia,Olmedo:2017lvt,Ashtekar:2018lag,Ashtekar:2018cay}, which partially alleviate the issues mentioned above.
\item \textit{$\bar{\mu}$-type scheme}

The polymerization scale in $\bar{\mu}$-type scheme is chosen to be a function of the phase space \cite{Boehmer:2007ket,Joe:2014tca,Chiou:2008nm,Chiou:2008eg} such that the dependency on fiducial structures is removed. Particularly, in the improved $\bar{\mu}$ scheme with Choui's choice \cite{Chiou:2008nm,Chiou:2008eg}, the spacetime approaches sufficiently fast the Schwarzschild geometry at low curvatures, which cures the drawback of the $\mu_0$-scheme \cite{Gambini:2020nsf,Kelly:2020uwj,Husain:2022gwp,Han:2022rsx}.
\end{itemize}

Recently, following the idea of the anomaly-free polymerization in \cite{Alonso-Bardaji:2021tvy}, a novel covariant model of a spherically symmetric BH with holonomy correction is proposed in \cite{Alonso-Bardaji:2021yls, Alonso-Bardaji:2022ear}. The polymerization scale $\bar{\lambda}$ is a constant in this model, and it is related to a fundamental length scal $r_0$ by a constant of motion $m$. The resulting geometry corresponds to a singularity-free interior region and two asymptotically flat exterior regions of equal mass.

In this paper, we will mainly study the properties of the quasi-normal modes (QNMs) of a probe scalar field and a probe Maxwell field over this covariant polymer BH. As we all know, during the ringdown phase of binary system coalescence, the BH emits the gravitational waves (GWs) with typical discrete frequencies, i.e., quasi-normal frequencies (QNFs). According to \cite{Berti:2009kk}, QNFs encode decaying scales and damped oscillating frequencies . Certainly, quantum effects have the imprints in the QNM spectra, which are expected to be detected in GW observations. Also, conversely, GW detection will serve as an important criterion for the correctness of candidate quantum gravity theories.

Our paper is organized as follows. In section \ref{sec-qsbh}, we present a brief discussion on the effective potentials of scalar and Maxwell fields over the covariant LQG-BH. Section \ref{sec_QNMs} is dedicated to the properties of the QNM spectra. Then, we further study the ringdown waveform in section \ref{sec-ring}. We present the conclusions and discussions in section \ref{sec-conclusion}. Appendixes \ref{app-wave-eq} and \ref{app-QNM-eikonal} present the detailed derivation of the wave equations and the QNMs in the eikonal limit.

\section{Effective quantum corrected Schwarzschild geometry}\label{sec-eqcsg}

In \cite{Alonso-Bardaji:2021yls, Alonso-Bardaji:2022ear}, the authors proposed a novel effective LQG corrected spherically symmetric black hole model with holonomy corrections that is covariant. In this section, we will present a brief review on this model.

\subsection{Effective quantum corrected Schwarzschild geometry}

In the framework of canonical GR depected by the Ashtekar-Barbero variables, a spherically symmetric model can be fully characterized by four dynamic variables: the two independent components of a densitized triad, $\widetilde{E}^x$ and $\widetilde{E}^{\varphi}$, along with their corresponding conjugate momenta, $\widetilde{K}_x$ and $\widetilde{K}_{\varphi}$. In this context, $x$ signifies the radial direction, while $\varphi$ represents the azimuthal angle. When incorporating holonomy corrections, the following canonical transformation is employed, as detailed in \cite{Alonso-Bardaji:2021yls, Alonso-Bardaji:2022ear}:
\begin{eqnarray}
	\widetilde{E}^x\rightarrow E^x\,,~~~~~\widetilde{K}_x\rightarrow K_x\,,~~~~~\widetilde{E}^{\varphi}\rightarrow\frac{E^{\varphi}}{\cos(\bar{\lambda} K_{\varphi})}\,,~~~~~\widetilde{K}_{\varphi}\rightarrow\frac{\sin(\bar{\lambda} K_{\varphi})}{\bar{\lambda}}\,.
	\label{CanT}
\end{eqnarray}
The parameter $\bar{\lambda}$, which can be conveniently taken as positive without loss of generality, is a dimensionless parameter inspired by holonomies. Notably, as $\bar{\lambda}$ approaches zero, the transformation is the identity, representing the limit where GR is recovered.

Under the transformation \eqref{CanT}, the diffeomorphism constraint remains unchanged, represented as 
\begin{eqnarray}
\mathcal{D}=-E^{x\prime}K_x+E^{\varphi}K^{\prime}_{\varphi}\,,
\label{DiffC}
\end{eqnarray}
where the prime denotes the derivative with respect to $x$. However, to ensure an algebra free of anomalies, it is necessary to perform a linear combination between the Hamiltonian constraint and the diffeomorphism constraint, as discussed in \cite{Alonso-Bardaji:2021tvy}. In addition, we also need regularize the poles $\cos(\bar{\lambda} K_{\varphi})=0$. Taking these considerations into account, a deformed Hamiltonian constraint can be constructed as \cite{Alonso-Bardaji:2021yls, Alonso-Bardaji:2022ear}:
\begin{eqnarray}
	\mathcal{H}&& =  -\frac{E^{\varphi}}{2 \sqrt{E^{x}} \sqrt{1+\bar{\lambda}^{2}}}\left(1+\frac{\sin ^{2}\left(\bar{\lambda} K_{\varphi}\right)}{\bar{\lambda}^{2}}\right)-\sqrt{E^{x}} K_{x} \frac{\sin \left(2 \bar{\lambda} K_{\varphi}\right)}{\bar{\lambda} \sqrt{1+\bar{\lambda}^{2}}}\left(1+\left(\frac{\bar{\lambda} E^{x \prime}}{2 E^{\varphi}}\right)^{2}\right) 
	\nonumber
	\\
	&& +\frac{\cos ^{2}\left(\bar{\lambda} K_{\varphi}\right)}{2 \sqrt{1+\bar{\lambda}^{2}}}\left(\frac{E^{x \prime}}{2 E^{\varphi}}\left(\sqrt{E^{x}}\right)^{\prime}+\sqrt{E^{x}}\left(\frac{E^{x \prime}}{E^{\varphi}}\right)^{\prime}\right)\,,\label{Hamiltonian-C}
\end{eqnarray}
along with its smeared form $H[f]:=\int f\mathcal{H}dx$. For more comprehensive details, please refer to \cite{Alonso-Bardaji:2021yls, Alonso-Bardaji:2022ear}.

By solving the system's equations of motion, which are derived from the Poisson brackets of various variables with the Hamiltonian, we can obtain the following explicitly spherically symmetric metric:
\begin{eqnarray}
ds^2=-N(t,x)^2dt^2+\Big(1-\frac{r_0}{\sqrt{E^x(t,x)}}\Big)^{-1}\frac{E^{\varphi}(t,x)^2}{E^x(t,x)}\Big(dx+N^x(t,x)dt\Big)^2+E^x(t,x)d\Omega^2\,,
\label{metricv0}
\end{eqnarray}
where $N$ and $N^x$ represent the lapse and shift functions, respectively. The parameter $r_0$ represents a new length scale and is determined by the expression:
\begin{eqnarray}
	r_0=2m\frac{\bar{\lambda}^2}{1+\bar{\lambda}^2}\,,
	\label{r0}
\end{eqnarray}
where $m$ stands as a constant of motion. Notably, the length scale $r_0$ sets a minimum area $r_0^2$ for this model \cite{Alonso-Bardaji:2021yls, Alonso-Bardaji:2022ear}. In contrast to its classical counterpart, it incorporates the term $1-r_0/\sqrt{E^x}$.

\begin{figure}[H]
	\center{
		\includegraphics[scale=0.7]{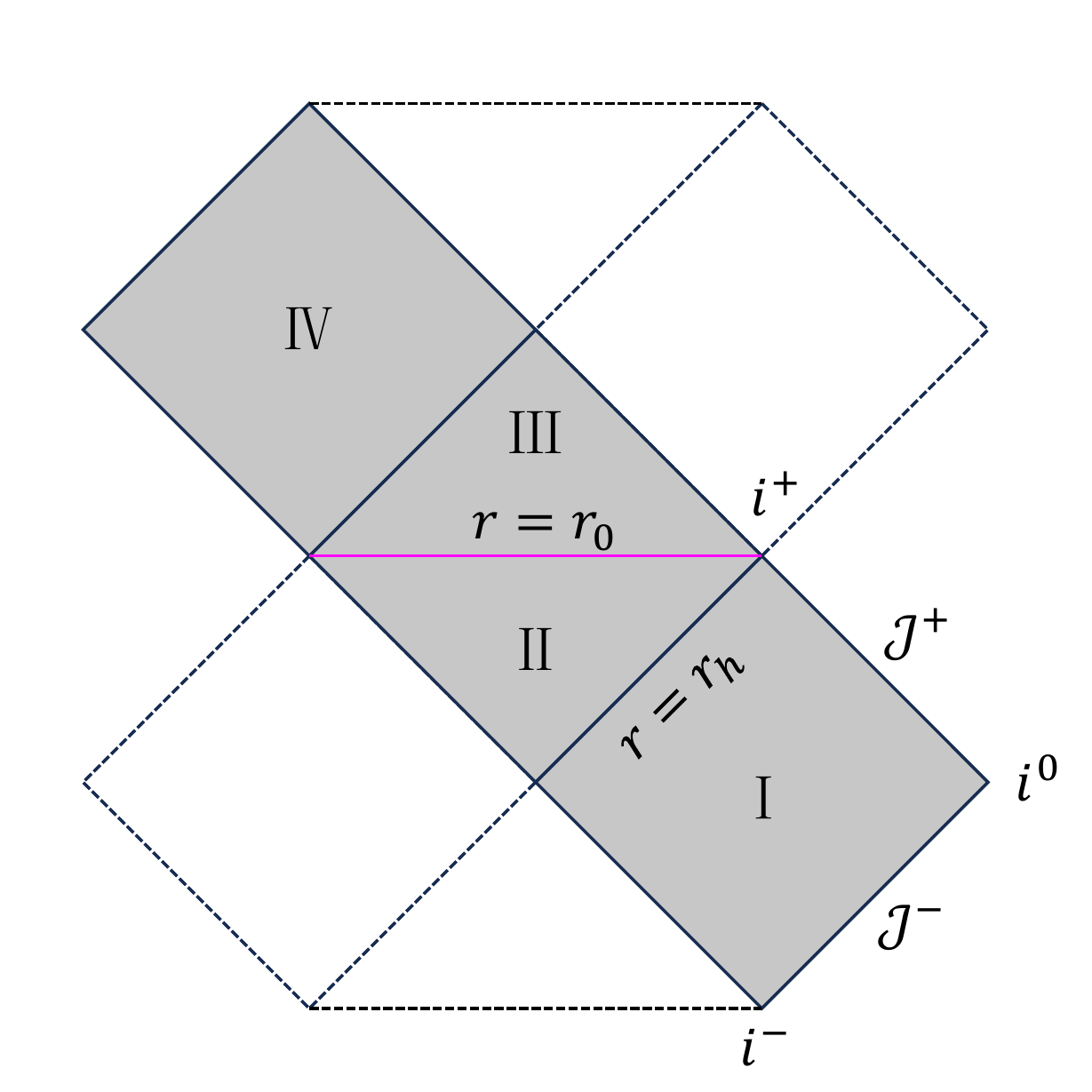}
		\caption{Penrose diagram with maximal analytical extension of the quantum-corrected spacetime.}
		\label{Penrose}
	}
\end{figure}

We would like to  emphasize that different gauge choices will result in distinct charts and their corresponding line elements for the same metric, ultimately yielding a consistent spacetime solution. The region containing the BH along with an asymptotically flat region can be described using the coordinate system ${t, x} = {t, r}$, in conjunction with the spherical metric ${\theta, \varphi}$. This quantum-corrected spacetime also possesses a maximal analytical extension \cite{Alonso-Bardaji:2021yls, Alonso-Bardaji:2022ear}, as depicted in the Penrose diagram shown in Fig.\ref{Penrose}. In this diagram:
\begin{itemize}
	\item Region I corresponds to the asymptotically flat region with $r \in (r_h, \infty)$. This region includes the usual conformal infinities, namely, the timelike infinities denoted as $i^{-}$ and $i^{+}$, the null infinities referred to as $\mathcal{J}^{-}$ and $\mathcal{J}^{+}$, as well as the spatial infinity marked as $i^0$.
	\item Region II is the BH region with $r \in (r_0, r_h)$. It is evident that the hypersurface $r=r_0$ is a transition surface between the BH and the WH regions.
	\item Regions III and IV are regions that cannot be covered by the coordinate system $(t, r, \theta, \varphi)$; they correspond to the white hole (WH) region and another asymptotically flat region, respectively.
\end{itemize}
The regions with dashed contours at the bottom and top are duplicates of the structure in the middle.

In this paper, our primary focus is the study of the QNMs for the BH part, thus limiting our attention to the exterior region of this quantum-corrected spacetime, as described by the following metric \cite{Alonso-Bardaji:2021yls, Alonso-Bardaji:2022ear}
\begin{eqnarray} \label{metric}
	&&
	ds^2=-f(r)dt^2+\frac{1}{g(r)f(r)}dr^2+r^2d\Omega^2\,,
	\nonumber
	\\
	&&
	f(r)=1-\frac{2m}{r}\,,\,\,\,\,\,\,\, g(r)=1-\frac{r_0}{r}\,,
\end{eqnarray}
where $r\in (r_h,\infty)$ with $r_h=2m$. Notably, in the limit as $\bar{\lambda}\rightarrow 0$, this new length scale tends to zero, i.e., $r_0=0$, thereby restoring the classical Schwarzschild BH. Furthermore, as we move to the low curvature regions, the quantum gravity effects die off.

\subsection{Mass and energy}

This section provides provide a concise overview of the typical geometric definitions of mass and energy applied to this solution. This discussion aids in gaining a thorough understanding of the model's parameters. A straightforward calculation reveals that the Komar mass is a function of $r$ and is expressed as follows \cite{Alonso-Bardaji:2022ear}: 
\begin{eqnarray}
M_K(r)=m\sqrt{1-\frac{r_0}{r}}\,.
\end{eqnarray}
It is readily apparent that the Komar mass approaches the constant of motion $m$ as it tends to infinity. The dependence of the Komar mass on $r$ is a consequence of the non-zero Ricci tensor.

The Hawking mass, equivalently, the Misner-Sharp mass, also exhibits a dependence on $r$ and can be expressed as follows \cite{Alonso-Bardaji:2022ear}:
\begin{eqnarray}
	M_H(r)=m+\frac{r_0}{2}-\frac{mr_0}{r}\,.
\end{eqnarray}
This quantity is consistently positive and, notably, coincides with $m$ solely at the horizon. It's evident that the presence of a non-zero Ricci tensor exerts distinct influences on the Komar and Hawking masses.

Another important quantity is the ADM mass. The ADM mass on the hypersurfaces $\Sigma_t$ can be worked out as follows \cite{Alonso-Bardaji:2021yls, Alonso-Bardaji:2022ear}:
\begin{eqnarray}
	M^t_{ADM}=m+\frac{r_0}{2}\,.
	\label{MADMt}
\end{eqnarray}
It's evident that the ADM mass is a geometric invariant and converges to the Hawking mass in the limit of infinity. While the ADM mass on any hypersurfaces $\Sigma_{\tau}$ is given by:
\begin{eqnarray}
	M^{\tau}_{ADM}=\frac{r_0}{2}\,.
	\label{MADMtau}
\end{eqnarray}
This result is characterizd by the parameter $r_0$ and recovers the one in the GR limit.

Additionally, we can calculate the Geroch energy. It is found that the Geroch energy on the hypersurface $\Sigma_t$ coincide with the Hawking energy. It is noteworthy that the Geroch energy on the hypersurface $\Sigma_t$ coincides with the Hawking energy. Meanwhile, the Geroch energy on $\Sigma_{\tau}$ equals the ADM mass of $\Sigma_{\tau}$, as expressed by:
\begin{eqnarray}
	E_G^{\tau}(r)=\frac{r_0}{2}\,.
\end{eqnarray}
An important point to highlight is that the Geroch mass is a quasi-local quantity.

We are also interested in the surface gravity $\kappa$, which is given by:
\begin{eqnarray}
	\kappa=\frac{1}{4 m}\sqrt{1-\frac{r_0}{2m}}\,.
	\label{kappa}
\end{eqnarray}
This fulfills the typical relation $|\kappa|=r^{-2}M_K|_{r=2m}$. It's worth noting that in the limit where $r_0\rightarrow 2m$, the surface gravity becomes zero, akin to the extremal Reissner-Nordstr\"{o}m spacetime. The existence of a minimum area results in the surface gravity being smaller than that of a Schwarzschild BH with mass $m$.

\section{Scalar field and Maxwell field over the LQG-BH}\label{sec-qsbh}

\begin{figure}[H]
	\center{
		\includegraphics[scale=0.85]{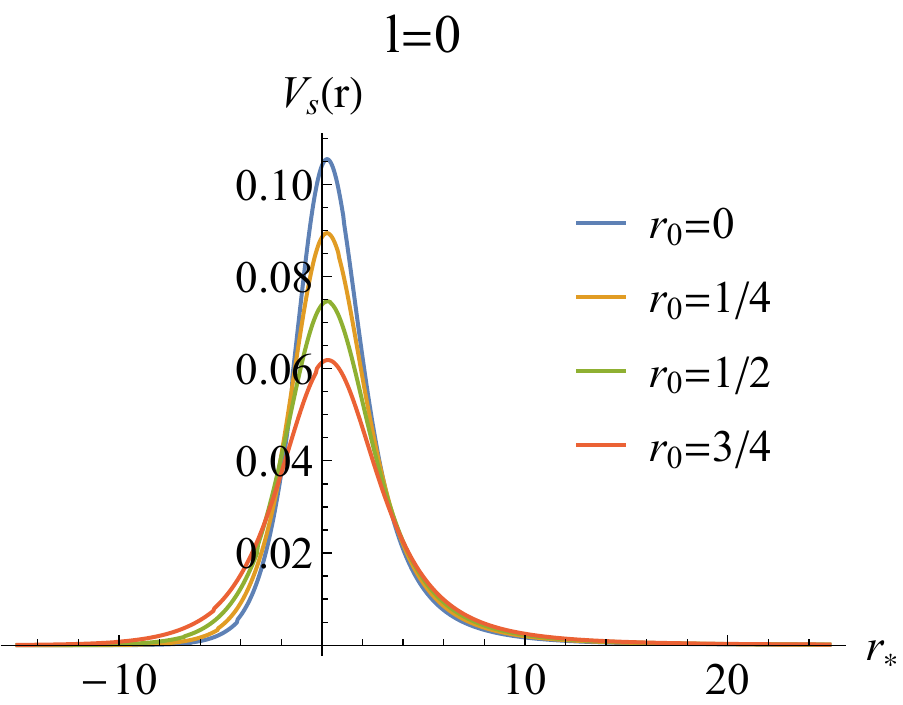}\hspace{0.2cm}
		\includegraphics[scale=0.85]{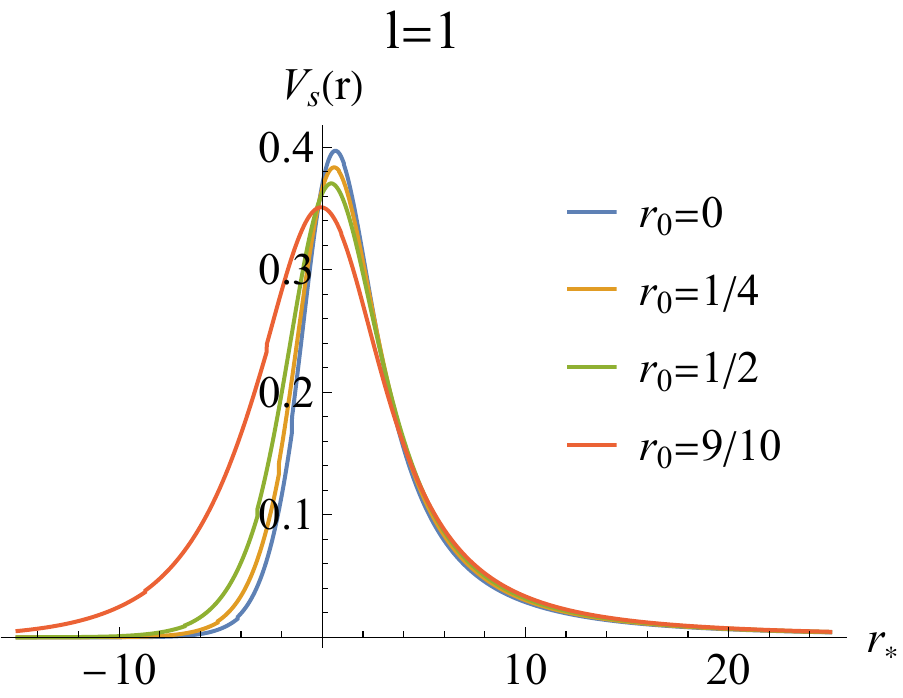}
		\caption{The effective potentials $V_{s}(r_*)$ of the scalar field for different $r_0$ with fixed $l$.}
		\label{Veff_scalar}
	}
\end{figure}

\begin{figure}[H]
	\center{
		\includegraphics[scale=0.65]{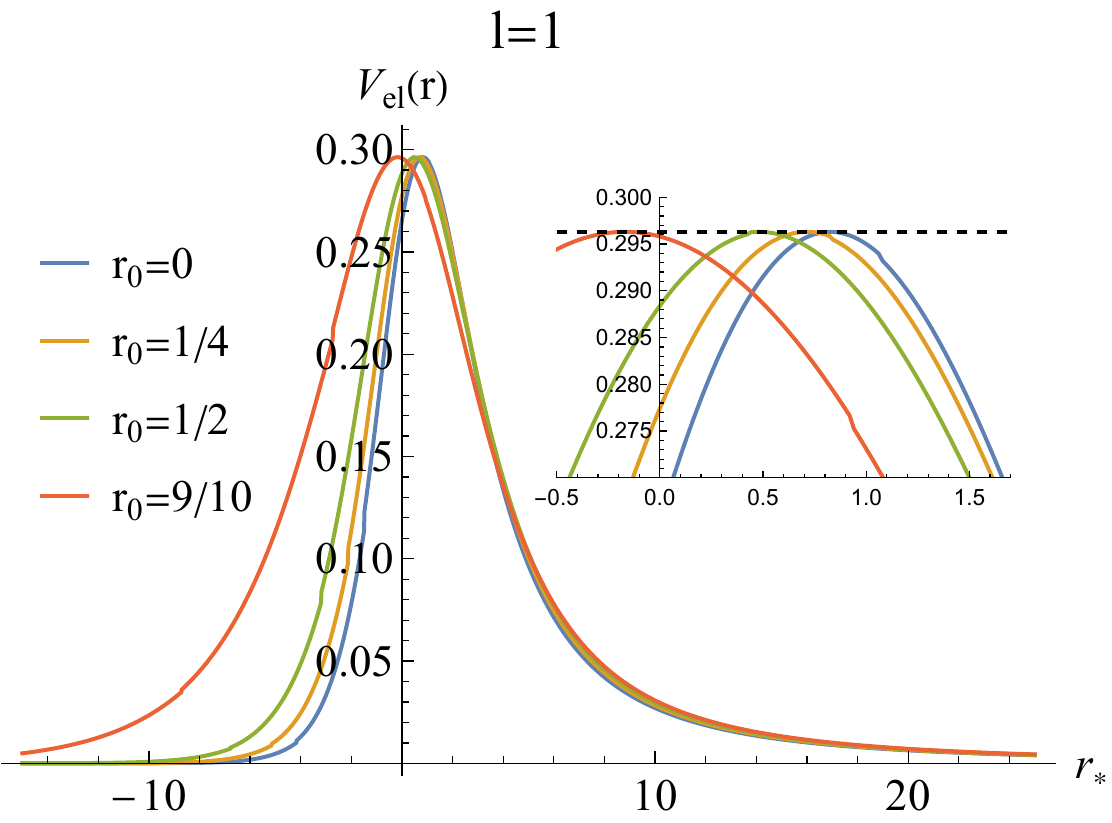}\hspace{0.2cm}
	    \includegraphics[scale=0.65]{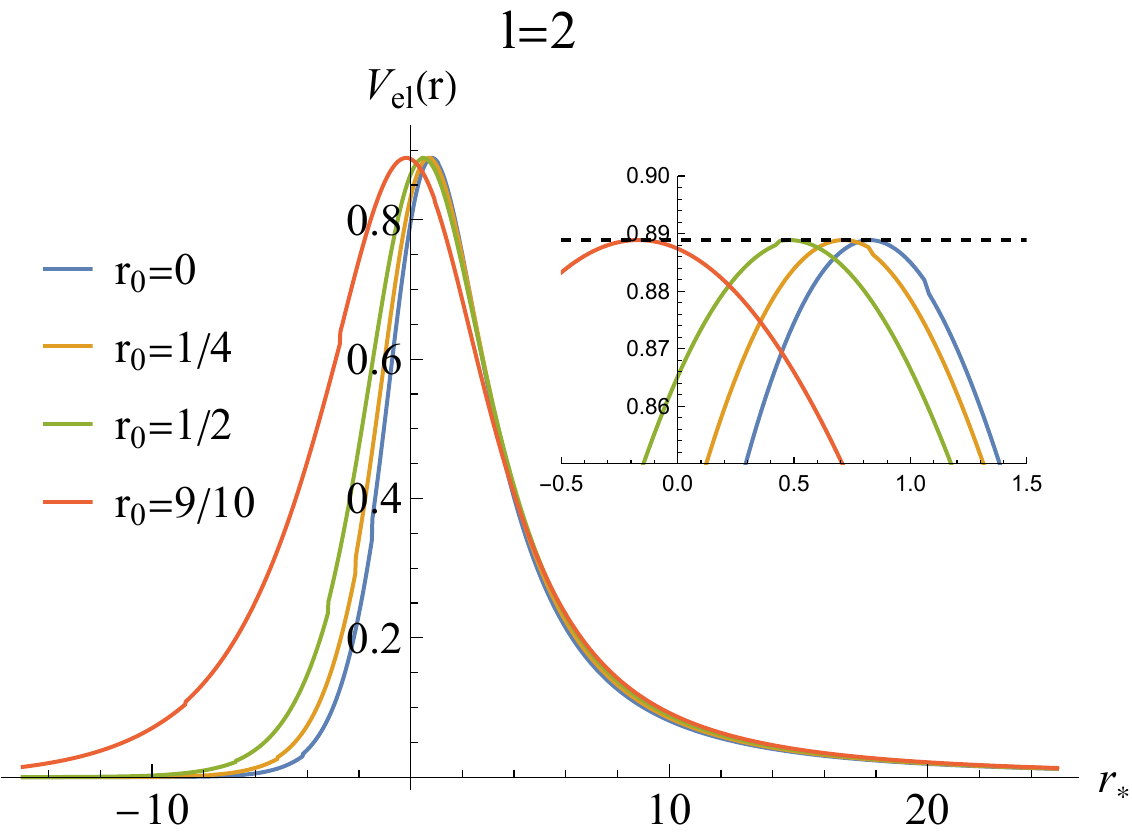}
		\caption{The effective potentials $V_{el}(r_*)$ of the electromagnetic field for different $r_0$ with fixed $l$.}
		\label{Veff_electro}
	}
\end{figure}

We focus on the perturbations of the massless scalar field $\Phi$ and electromagnetic field $A_\mu$ over this LQG black hole and study their response. Notice that in the following, we shall set set $m=1/2$ without loss of generality, which leads to the horizon located at $r_h=1$. We write down the covariant equations for the test scalar field and electromagnetic field as follows:
\begin{eqnarray}
&&
\label{scalar_eq}
\frac{1}{\sqrt{-g}}(g^{\mu\nu}\sqrt{-g}\Phi_\mu),_\nu=0\,, \ \
\\
&&
\label{electro_eq}
\frac{1}{\sqrt{-g}}(g^{\alpha\mu}g^{\sigma\nu}\sqrt{-g} F_{\alpha \sigma}),_{\nu}=0\,,
\end{eqnarray}
where $F_{\alpha \sigma}= \partial_\alpha A_\sigma- \partial_\sigma A_\alpha$ is the field strength of the Maxwell field. After the separation of variables, the aforementioned equations can be packaged into the Schr\"{o}dinger-like form (for more details, see Appendix \ref{app-wave-eq})
\begin{eqnarray}\label{Sch_like_eq}
\frac{\partial ^2 \Psi}{\partial r_{*}^2}+(\omega^2 -V_{eff}) \Psi=0\,,
\end{eqnarray}
where $r_*$ is the tortoise coordinate and $V_{eff}$ is the effective potentials:
\begin{eqnarray}\label{V_eff}
V_{eff}=f(r)\frac{l(l+1)}{r^2}+\frac{1-s}{r}\frac{d}{dr_*}f(r)\sqrt{g(r)}\,,
\end{eqnarray}
with $l$ being the angular quantum numbers. $s=0$ and $s=1$ correspond to the scalar field and electromagnetic field, respectively.
Fig.\ref{Veff_scalar} and Fig.\ref{Veff_electro} demonstrate the effective potentials as a function of $r_*$ for scalar and electromagnetic fields with different $l$ and $r_0$. It is found that both effective potentials are positive, indicating the LQG black hole is stable under scalar and electromagnetic perturbations.
Furthermore, we would like to compare the differences in effective potentials between scalar and electromagnetic fields.
It is easy to find that for the electromagnetic field ($s=1$), the second term in Eq.\eqref{V_eff} vanishes, such that all the peaks of the effective potentials $V_{el}$ have the same height for different $r_0$ (see Fig.\ref{Veff_electro}). However, for the scalar field, i.e., $s=0$, the second term in Eq.\eqref{V_eff} survives and the height of the effective potential $V_s$ depends on $r_0$. In particular, with increasing $r_0$, the height of $V_s$ decreases (Fig.\ref{Veff_scalar}). The shape of the effective potentials shall definitely results in different properties of the QNMs.


\section{Quasi-normal modes}\label{sec_QNMs}

In this section, we investigate the QNMs spectra and specially focus on the effects from quantum gravity corrections. The nature of determining the QNMs is to solve the eigenvalue problem. To this end, we will impose a purely outgoing wave at infinity and purely ingoing wave at the horizon:
\begin{eqnarray}\label{boundary_cond}
	&&
\text{horizon:}\ \ \partial_{t}\Psi-\partial_{r_*}\Psi=0 \red{,}\ \ \nonumber \\
    &&
\text{infinity:}\ \ \partial_{t}\Psi+\partial_{r_*}\Psi=0 \red{.}
\end{eqnarray}
By solving Eq.\eqref{Sch_like_eq} with the aforementioned boundary conditions, numerous techniques have been developed to determining the QNMs spectra, such as the WKB method \cite{1985ApJ-291L33S,Iyer:1986np,Guinn:1989bn,Konoplya:2004ip,Konoplya:2003ii,Matyjasek:2017psv}, Horowitz-Hubeny method \cite{Horowitz:1999jd}, continued fraction method \cite{Leaver:1985ax}, asymptotic iteration method \cite{Ciftci:2005xn,Cho:2009cj,Cho:2011sf}, pseudo-spectral method \cite{Boyd:Chebyshev,Jansen:2017oag}, and so on. In this paper, we will solve the eigenvalue problem using the pseudo-spectral method. For more applications of pseudo-spectral method in determining the QNMs in the black hole physics, we can refer to \cite{Wu:2018vlj,Fu:2018yqx,Fu:2022cul,Xiong:2021cth,Liu:2021fzr,Liu:2021zmi,Jaramillo:2020tuu,Jaramillo:2021tmt,Destounis:2021lum,Mamani:2022akq} and references therein. It is convenient to work in the Eddington-Finkelstein coordinate, which makes the wave equation \eqref{Sch_like_eq} is linear in the frequency. To achieve this goal, it is direct to make
a transformation as
\begin{eqnarray}
r\to 1/u \ \ \text{and} \ \   \Psi=e^{- i \omega r_*(u)}\psi\,.
\end{eqnarray}
Then, the wave equation \eqref{Sch_like_eq} turns into the following form:
\begin{eqnarray}\label{eq_EF}
&&
\psi''(u)+\left(\frac{f'(u)}{f(u)}+\frac{g'(u)}{2g(u)}+\frac{2i\omega}{u^2f(u)\sqrt{g(u)}}\right)\psi'(u) \nonumber \\
&&
-\frac{1}{u}\left(\frac{2i\omega}{u^2f(u)\sqrt{g(u)}}+\frac{V_{eff}(u)}{u^3f(u)^2g(u)}+\frac{f'(u)}{f(u)}+\frac{g'(u)}{2g(u)}\right)\psi(u)=0
\red{.}\end{eqnarray} 
Combining with the boundary conditions \eqref{boundary_cond}, one can solve Eq.\eqref{eq_EF} by the pseudo-spectral method.

\begin{figure}[H]
	\center{
		\includegraphics[scale=0.8]{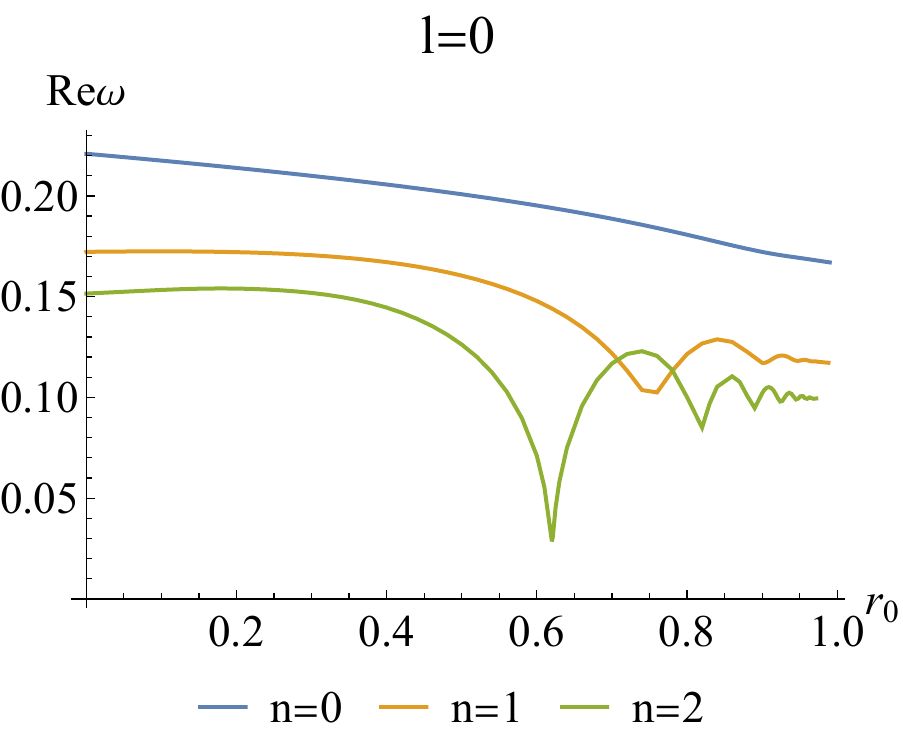}\hspace{0.1cm}
		\includegraphics[scale=0.85]{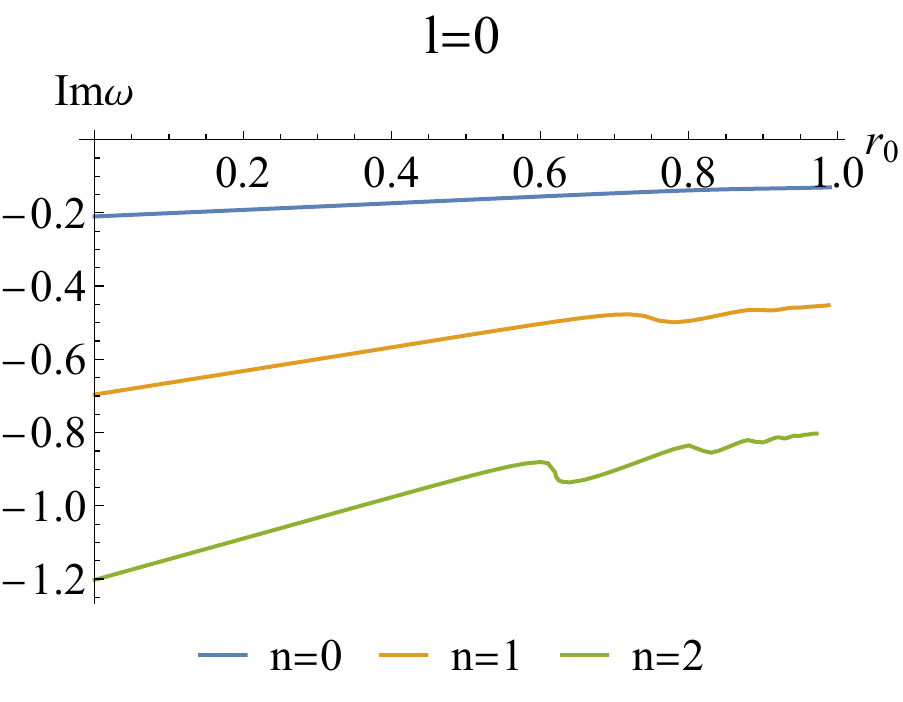}\hspace{0.1cm}
		\includegraphics[scale=0.8]{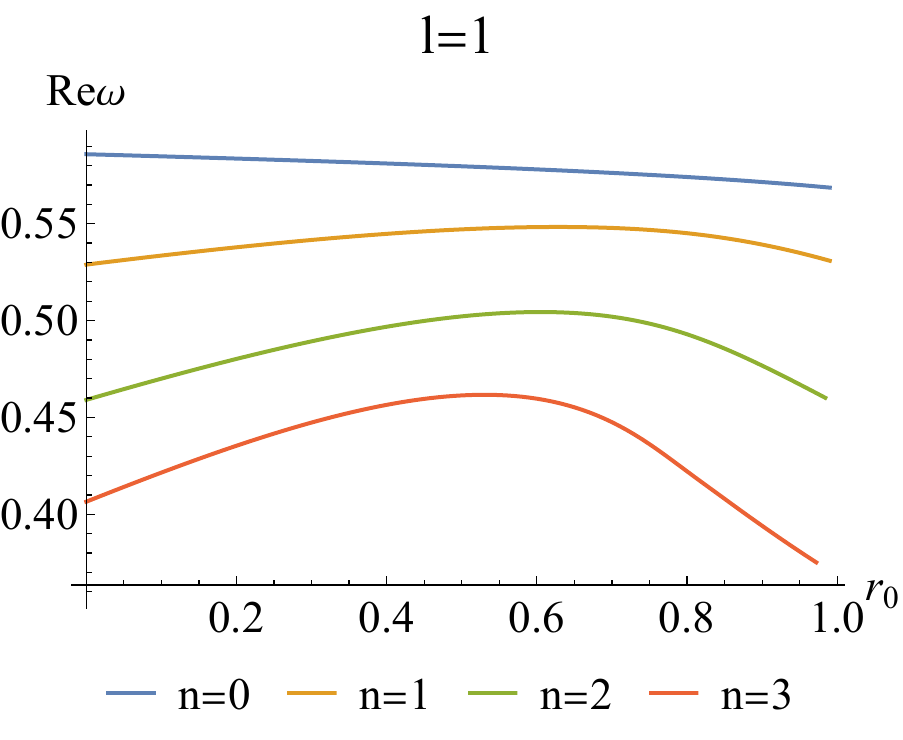}\hspace{0.1cm}
		\includegraphics[scale=0.85]{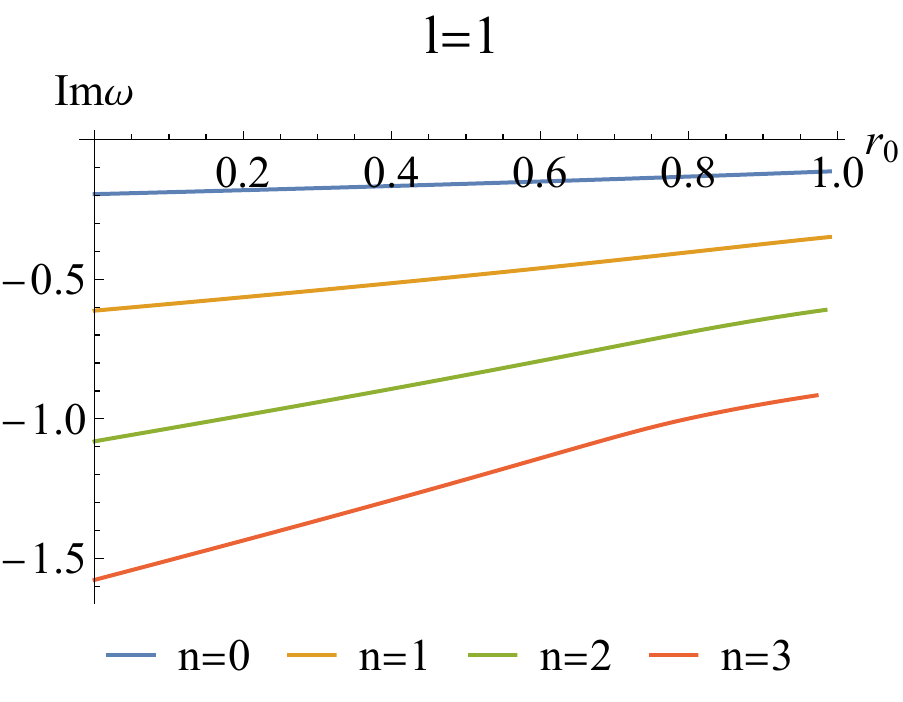}\hspace{0.1cm}
		\caption{QNFs as a function of $r_0$ for the scalar field perturbation.}
		\label{QNMs_scalar}
	}
\end{figure}

\begin{figure}[H]
	\center{
		\includegraphics[scale=0.8]{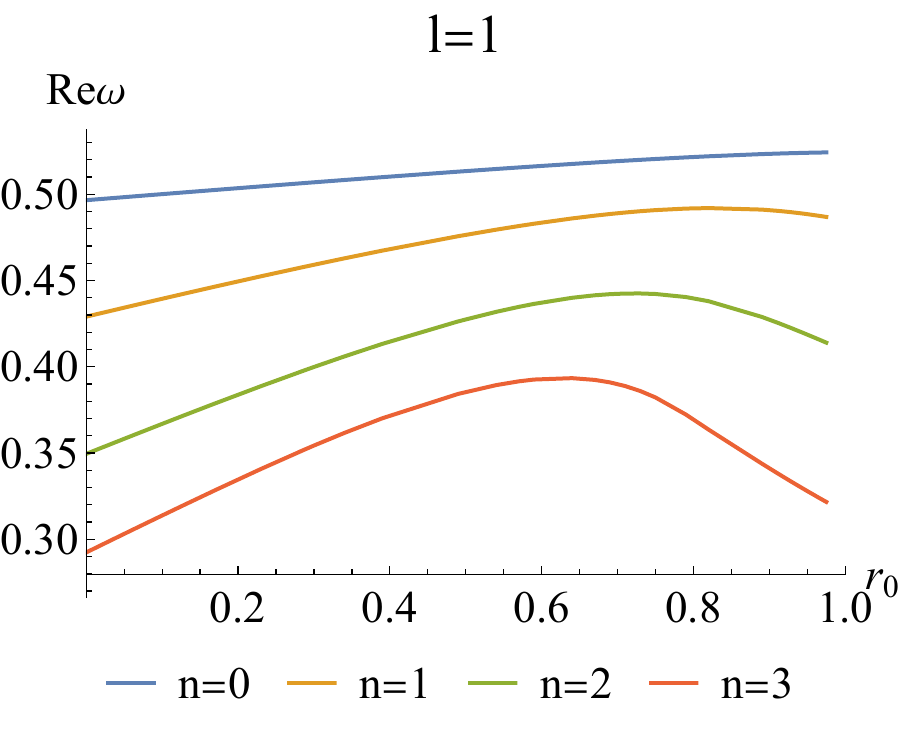}\hspace{0.1cm}
		\includegraphics[scale=0.85]{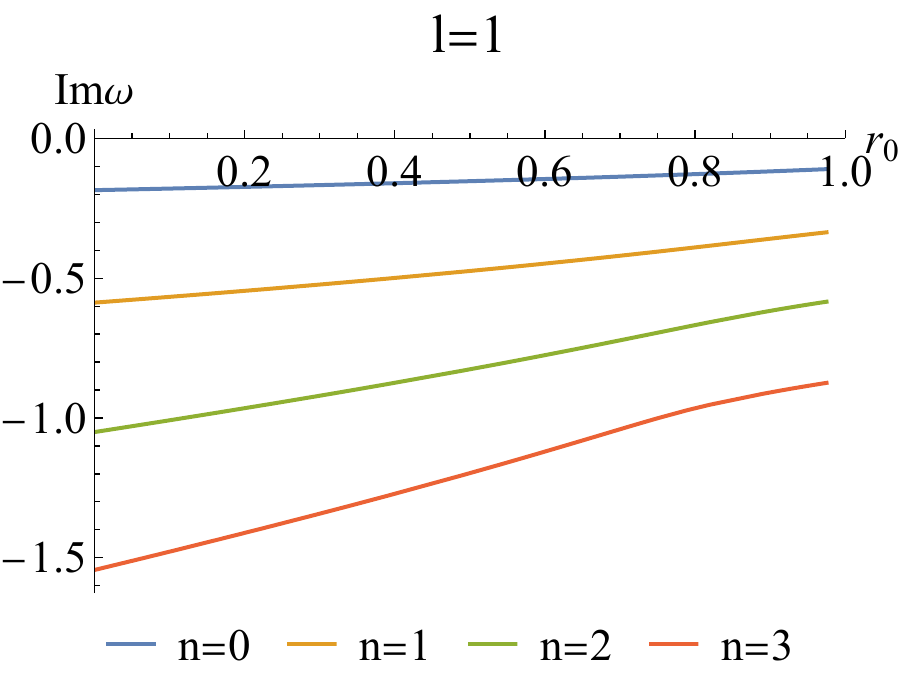}\hspace{0.1cm}
		\includegraphics[scale=0.8]{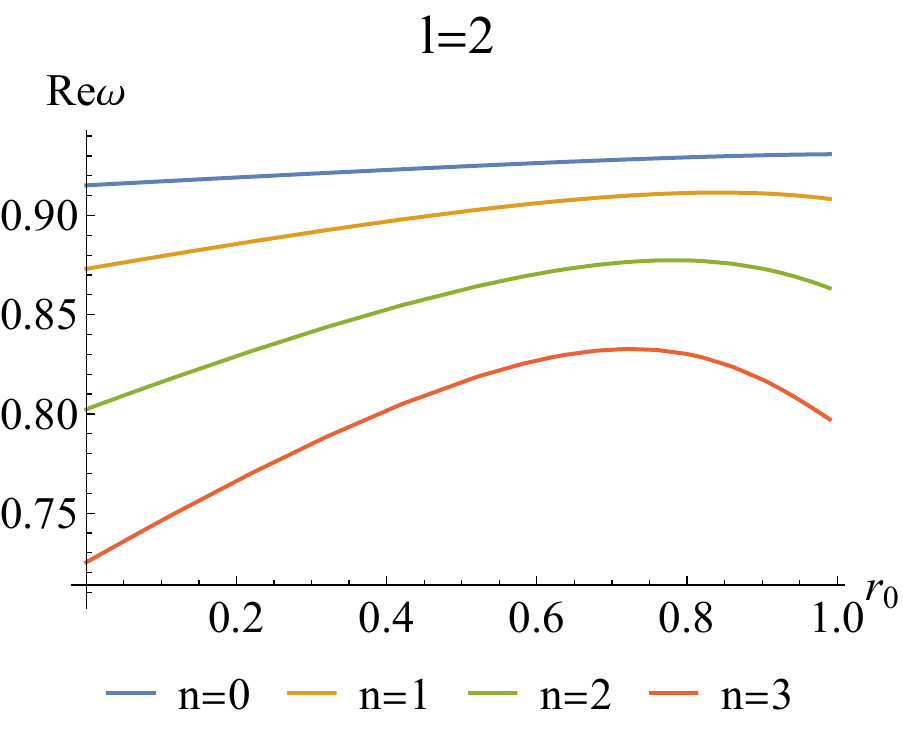}\hspace{0.1cm}
		\includegraphics[scale=0.85]{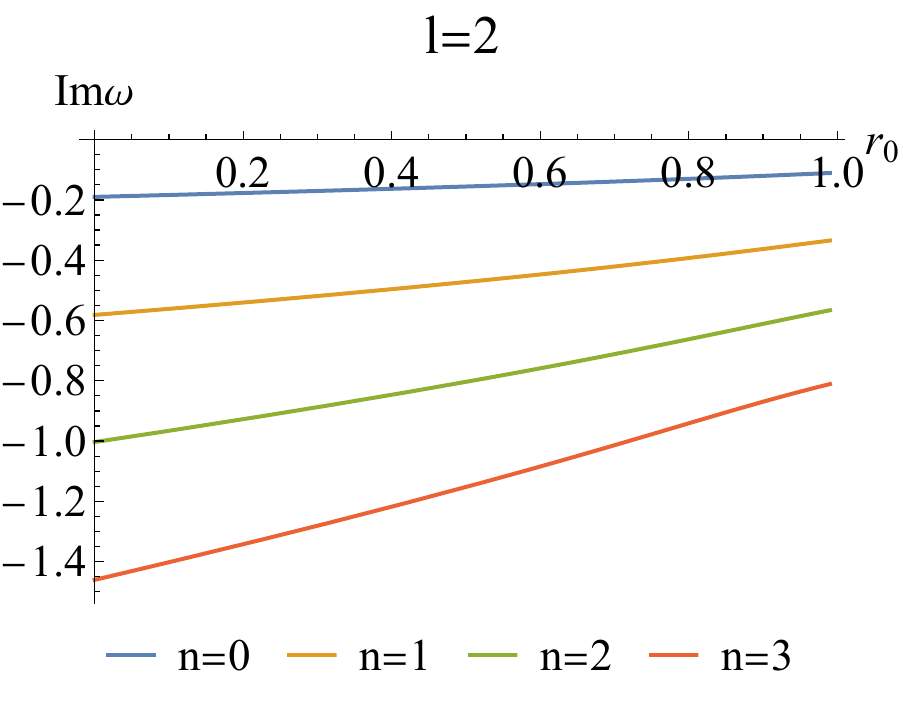}\hspace{0.1cm}
		\caption{QNFs as a function of $r_0$ for the electromagnetic field perturbation.}
		\label{QNMs_electro}
	}
\end{figure}

Now, we evaluate the QNM spectra for various values of the free parameter $r_0$ to explore the LQG effects on these spectra, as well as the differences between them and those of the Schwarzschild BH ($r_0=0$). Fig.\ref{QNMs_scalar} and Fig.\ref{QNMs_electro} show the QNFs as functions of $r_0$ for both the scalar field and electromagnetic field, respectively, showcasing multiple overtone numbers. We also provide the values of the QNFs for some specific parameter $r_0$ in Tables \ref{TABLE-1}, \ref{TABLE-2}, \ref{TABLE-3} and \ref{TABLE-4}. We firstly summarize the properties of the fundamental modes ($n=0$) as follows.
\begin{itemize}
	\item For scalar field, the real parts of the QNF, $Re\omega$ decreases with increasing $r_0$ (left plots in Fig.\ref{QNMs_scalar}). It means that the LQG effect reduces the oscillations in comparison to the Schwarzschild black hole. By contrast, $Re\omega$ of the electromagnetic field exhibits an inverse tendency. That is, when $r_0$ increases, so does $Re\omega$. As a result, the LQG effect produces stronger oscillations in the electromagnetic field than that of the Schwarzschild black hole.
	\item Whether for scalar or electromagnetic field, the imaginary part of QNF $Im\omega$ always lives in the lower half-plane, and their absolute values are less than that of the Schwarzschild black hole. Therefore, the system is stable in the presence of scalar or electromagnetic field perturbations, and the LQG effect results in slower decaying modes.
	\item When we fix $r_0$, the scalar field has larger absolute values of $Re\omega$ or $Im\omega$ than the electromagnetic field (see Figs.\ref{QNMs_scalar} and \ref{QNMs_electro}). It indicates that in comparison to the electromagnetic field, the system under scalar field perturbation enjoys faster decaying modes with greater oscillations.
\end{itemize}

\begin{table*}[htp]
	\centering
	\begin{tabular}{c|cccc}
		\hline n & $\omega$  ($r_0=0$)  & $\omega$  ($r_0=1/100$)& $\omega$  ($r_0=1/2$)  & $\omega$  ($r_0=9/10$)   \\
		\hline
		0 & 0.220910-0.209792i & 0.220578-0.208911i & 0.200799-0.164527i & 0.172186-0.133779i 
		\\
		1 & 0.172223-0.696106i & 0.172273-0.692920i & 0.160438-0.534669i & 0.117132-0.465373i 
		\\
		2 & 0.151564-1.202642i & 0.151650-1.196552i & 0.126266-0.921455i & 0.102060-0.826914i 
		\\
		3 & 0.142272-1.705216i & 0.141476-1.699571i & 0.076951-1.314823i & 0.098727-1.175255i 
		\\
		4 & 0.134739-2.211987i & 0.133115-2.201015i & 0.053109-1.808749i & 0.094052-1.519302i 
		\\
		5 & 0.129639-2.712112i & 0.129003-2.704325i & 0.098367-2.213623i & 0.083108-1.869088i 
		\\
		\hline
	\end{tabular}
	\caption{The QNM spectra for the scalar field perturbation for $l=0$ with different $n$ and $r_0$.\label{TABLE-1}}
\end{table*}

\begin{table*}[htp]
	\centering
	\begin{tabular}{c|cccc}
		\hline n & $\omega$  ($r_0=0$)  & $\omega$  ($r_0=1/100$)  & $\omega$   ($r_0=1/2$) &  $\omega$  ($r_0=9/10$) \\
		\hline
		0 & 0.585872-0.195320i & 0.585765-0.194632i &  0.579649-0.158265i & 0.571522-0.122834i
		\\
		1 & 0.528897-0.612515i & 0.529378-0.610151i &  0.547089-0.487735i & 0.539232-0.374106i
		\\
		2 & 0.459079-1.080267i & 0.460200-1.075690i &  0.502191-0.843179i & 0.476752-0.643355i
		\\
		3 & 0.406517-1.576596i & 0.408055-1.569596i &  0.461392-1.216781i & 0.394074-0.946941i
		\\
		4 & 0.370218-2.081524i & 0.372040-2.072076i &  0.426854-1.597881i & 0.317781-1.289254i
		\\
		5 & 0.344154-2.588236i & 0.346193-2.576352i &  0.395550-1.981462i & 0.263692-1.650819i
		\\
		\hline
	\end{tabular}
	\caption{The QNM spectra for the scalar field perturbation for $l=1$ with different $n$ and $r_0$.\label{TABLE-2}}
\end{table*}

\begin{figure}[H]
	\center{
		\includegraphics[scale=0.8]{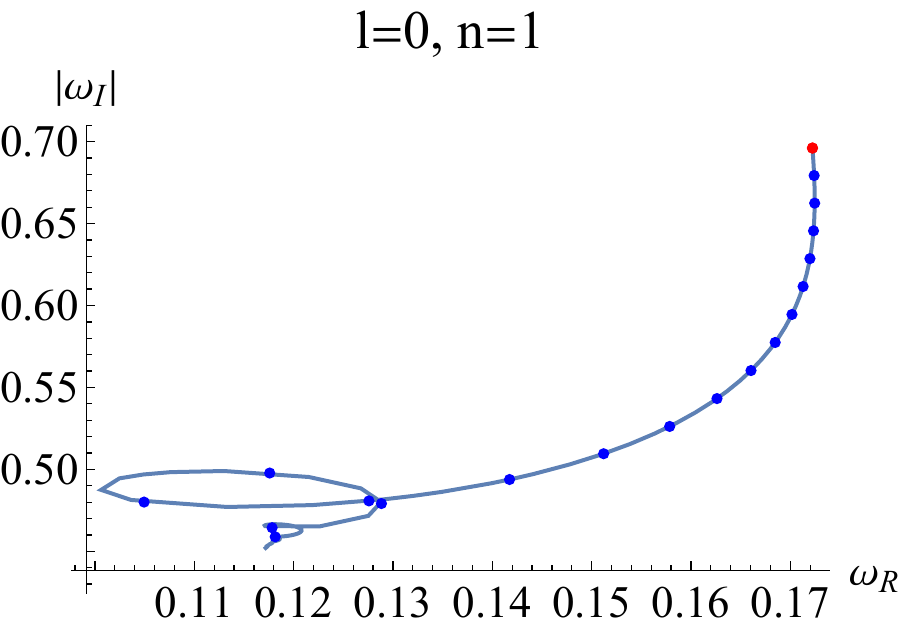}\hspace{0.1cm}
		\includegraphics[scale=0.85]{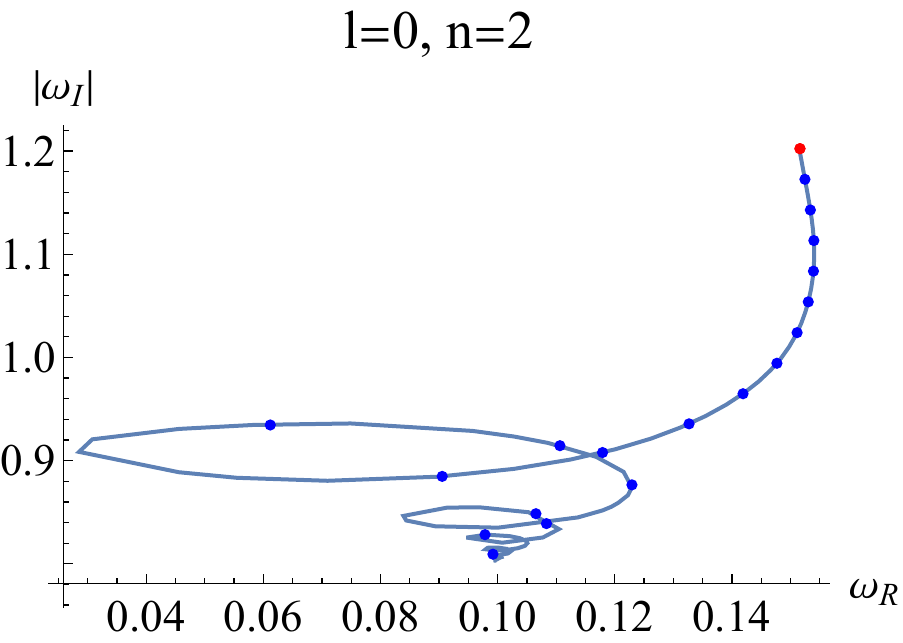}\hspace{0.1cm}
		\caption{QNFs of the scalar field perturbation with $n=1$ (left) and $n=2$ (right) for $r_0$ range from $0$ to $0.99$. The red point represent the $r_0=0$ (SS-BH) and the blue point correspond to $19r_0=1,2,...,18$. }
		\label{QNMs_scalar_n}
	}
\end{figure}

Next, we delve into an investigation of the properties of the QNM spectra with higher overtones. It is found that higher overtones exhibit some peculiar properties, differing significantly from those of the fundamental modes. We will now summarize their key properties.
\begin{itemize}
	\item For both the scalar field and the electromagnetic field, the first several overtones exhibit a substantially higher rate of deviation from their Schwarzschild values when compared to the fundamental mode, and this deviation rate increases with the overtone number. (see Figs.\ref{QNMs_scalar} and \ref{QNMs_electro}, also see Tables \ref{TABLE-1}, \ref{TABLE-2}, \ref{TABLE-3} and \ref{TABLE-4}). Especially, for the scalar field with $l=0$, the real oscillation frequency of the second overtone falls by more than six times its Schwarzschild limit, but the fundamental mode just slightly changes. This emergence of the outburst of overtones can be attributed to differences in the region near the event horizon between the Schwarzschild BH and the LQG corrected BH (see Figs.\ref{Veff_scalar} and \ref{Veff_electro}). Similar phenomena have been observed in alternative geometries beyond the Schwarzschild BH, including the Reissner-Nordstr\"{o}m BH, Bardeen BH and the higher-derivative gravity model \cite{Berti:2003zu,Konoplya:2022hll,Konoplya:2022pbc,Konoplya:2022iyn,Konoplya:2023ppx,Konoplya:2023ahd}.
	\item A remarkable phenomenon is seen wherein the QNFs of the scalar field with $l=0$ exhibit an oscillatory behavior when the quantum parameter $r_0$ increases significantly for higher overtones. The oscillations intensify with an increase in the overtone number. Similar oscillatory behaviors are also observed in Reissner-Nordstr\"{o}m  and Kerr BHs when these BHs approach the extremal case, as noted in \cite{Berti:2003zu}. We hypothesize that this oscillatory behavior may be linked to the extremal effect and, as such, warrants further investigation. Additionally, we suspect that similar oscillatory patterns may emerge in the case of a scalar field with high $l$ and in the context of the electromagnetic field when the overtone number is sufficiently high. Unfortunately, our current numerical techniques constrain us from calculating QNFs for higher overtone numbers at this time. We hope to address this issue in the near future.
	\item In Fig.\ref{QNMs_scalar_n}, the phase diagram $\omega_R$-$|\omega_I|$ is presented. The trajectory depicted in the phase diagram diverges from the Schwarzschild QNMs and spirals toward a stable point. This discovery aligns with the findings in \cite{Moreira:2023cxy}. This phenomenon actually reflects the previously mentioned oscillatory behavior and has also been observed in the context of the Reissner-Nordstr\"{o}m and Kerr BHs \cite{Berti:2003zu,Jing:2008an}.
\end{itemize}

\begin{table*}[htp]
	\centering
	\begin{tabular}{c|cccc}
		\hline n & $\omega$  ($r_0=0$)  & $\omega$  ($r_0=1/100$) & $\omega$  ($r_0=1/2$)  & $\omega$  ($r_0=9/10$) \\
		\hline
		0 & 0.496527-0.184975i & 0.496879-0.184421i & 0.513377-0.152855i & 0.523384-0.117771i \\
		1 & 0.429031-0.587335i & 0.430079-0.585304i & 0.476434-0.474099i & 0.490764-0.358757i \\
		2 & 0.349547-1.050375i & 0.351295-1.046207i & 0.427459-0.825862i & 0.427304-0.616791i \\
		3 & 0.292353-1.543818i & 0.294508-1.537311i & 0.385422-1.197325i & 0.341172-0.908634i \\
		4 & 0.253105-2.045090i & 0.255501-2.036119i & 0.351646-1.576036i & 0.259632-1.241057i \\
		5 & 0.224562-2.547950i & 0.227824-2.536499i & 0.322640-1.956797i & 0.199813-1.593456i \\
		\hline
	\end{tabular}
	\caption{The QNM spectra for the electromagnetic field perturbation for $l=1$ with different system parameters $n$ and $r_0$.\label{TABLE-3}}
\end{table*}
\begin{table*}[htp]
	\centering
	\begin{tabular}{c|cccc}
		\hline n & $\omega$  ($r_0=0$)  & $\omega$   ($r_0=1/100$) & $\omega$   ($r_0=1/2$) & $\omega$   ($r_0=9/10$)\\
		\hline
		0 & 0.915191-0.190009i & 0.915392-0.189392i & 0.924716-0.155637i & 0.930332-0.120325i \\
		1 & 0.873085-0.581420i & 0.873735-0.579444i & 0.901934-0.472399i & 0.911143-0.362898i \\
		2 & 0.802373-1.003175i & 0.803766-0.999483i & 0.862549-0.803573i & 0.873233-0.611234i \\
		3 & 0.725190-1.460397i & 0.727327-1.454593i & 0.816205-1.152343i & 0.817382-0.869439i \\
		4 & 0.657473-1.943219i & 0.660188-1.935080i & 0.770903-1.515796i & 0.745134-1.142754i \\
		5 & 0.602986-2.439431i & 0.606126-2.428886i & 0.730157-1.888692i & 0.661207-1.437692i \\
		\hline
	\end{tabular}
	\caption{The QNM spectra for the electromagnetic field perturbation for $l=2$ with different system parameters $n$ and $r_0$.\label{TABLE-4}}
\end{table*}

\begin{figure}[H]
	\center{
\includegraphics[scale=0.8]{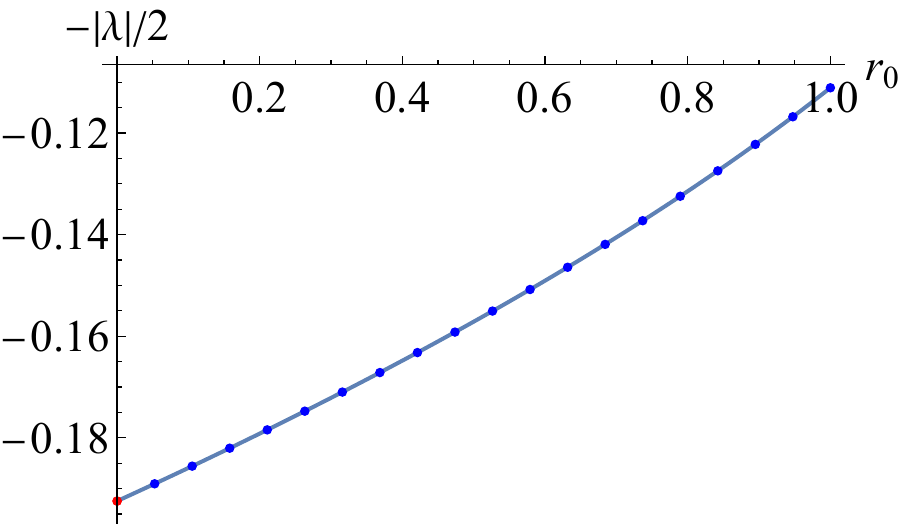}\hspace{0.1cm}
\includegraphics[scale=0.45]{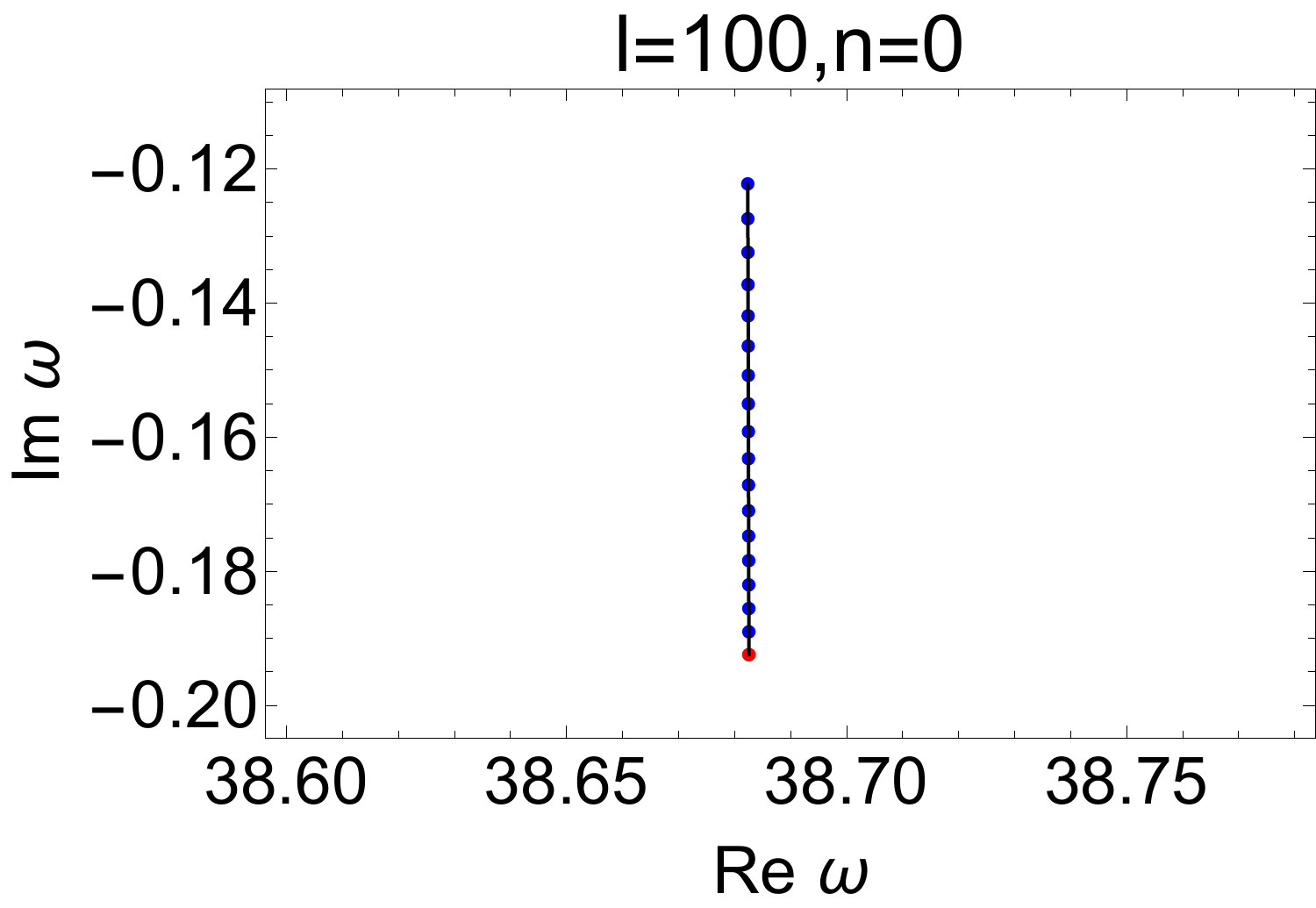}\hspace{0.1cm}
		\caption{Left plot: The Lyapunov exponent $\lambda$ as a function of the LQG corrected parameter $r_0$. Right plot: The QNFs for different $r_0$ for large $l$.  The red point represent the $r_0=0$ (SS-BH) and the blue point correspond to $19r_0=1,2,...,18$.}
		\label{Lyapunov-vs-r0}
	}
\end{figure}

Finally, we will discuss the properties of the QNMs in the eikonal limit ($l\rightarrow \infty$). In \cite{Cardoso:2008bp}, Cardoso $et$ $al$ have demonstrated that, in the eikonal limit, QNMs may be connected with the behavior of null particle trapped on the unstable circular geodesic of the spacetime, which have been validated in most static, spherically symmetric, asymptotically flat spacetime. The $Re\omega$ is determined by the angular velocity $\Omega_c$ at the unstable null geodesic \cite{Wei:2019jve,Jusufi:2019ltj,Guo:2020blq,Liu:2020ola,Ling:2021vgk}, whereas the $Im\omega$ is connected to the Lyapunov exponent $\lambda$ \cite{Bombelli:1991eg,Cornish:2003ig}. In the LQG-BH background, we can calculate the QNMs in the eikonal limit, which is given by
\begin{eqnarray}\label{QNMs_eikonal}
	\omega=\Omega_c l-i\Big(n+\frac{1}{2}\Big)|\lambda|\,.
\end{eqnarray}
For the detailed calculation, we can refer to Appendix \ref{app-QNM-eikonal}.
It is found that as SS-BH, the angular velocity $\Omega_c$ is completely determined by the black hole mass:
\begin{eqnarray}
	\Omega_c=\frac{1}{3\sqrt{3}m}\,.
\end{eqnarray}
Therefore, the $Re\omega$ is independent of the LQG parameter $r_0$.
While the Lyapunov exponent $\lambda$ is given by
\begin{eqnarray}
	\lambda=\sqrt{-\frac{r_c^2}{f(r_c)}\left(\frac{d^2}{dr_*^2}\frac{f(r)}{r^2}\right)\Big{|}_{r=rc}}\,,
\end{eqnarray}
where $r_c$ is the radius of the photon sphere. Obviously, the Lyapunov exponent is affected by the LQG correction. Left plot in Fig.\ref{Lyapunov-vs-r0} shows the Lyapunov exponent $\lambda$ as a function of $r_0$. We see that the Lyapunov exponent decreases with $r_0$ increasing. Correspondingly, the absolution value of $Im\omega$ is suppressed by the the LQG effect (see the right plot in Fig.\ref{Lyapunov-vs-r0}).

\begin{figure}[H]
	\center{
		\includegraphics[scale=0.45]{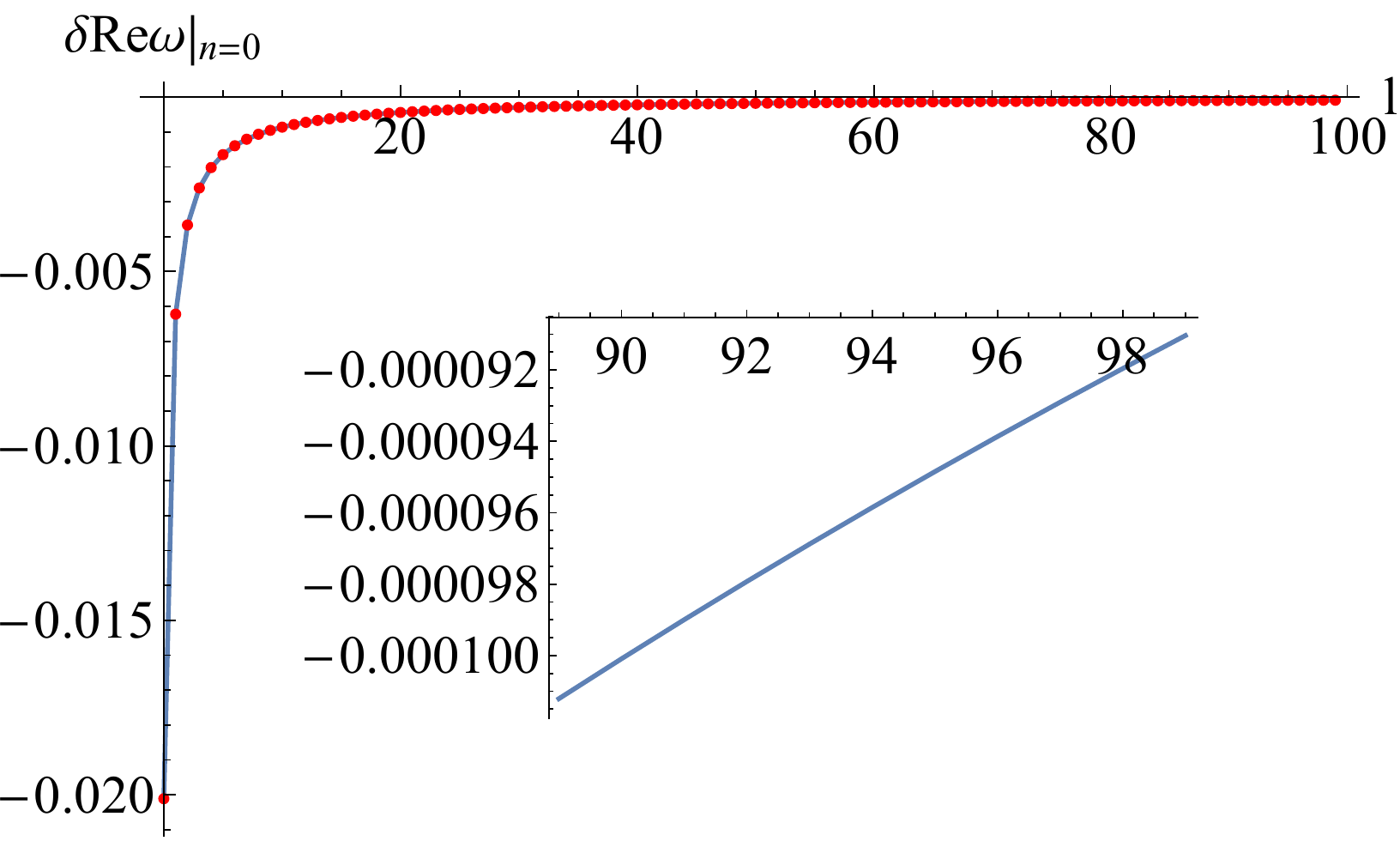}\hspace{0.1cm}
		\includegraphics[scale=0.45]{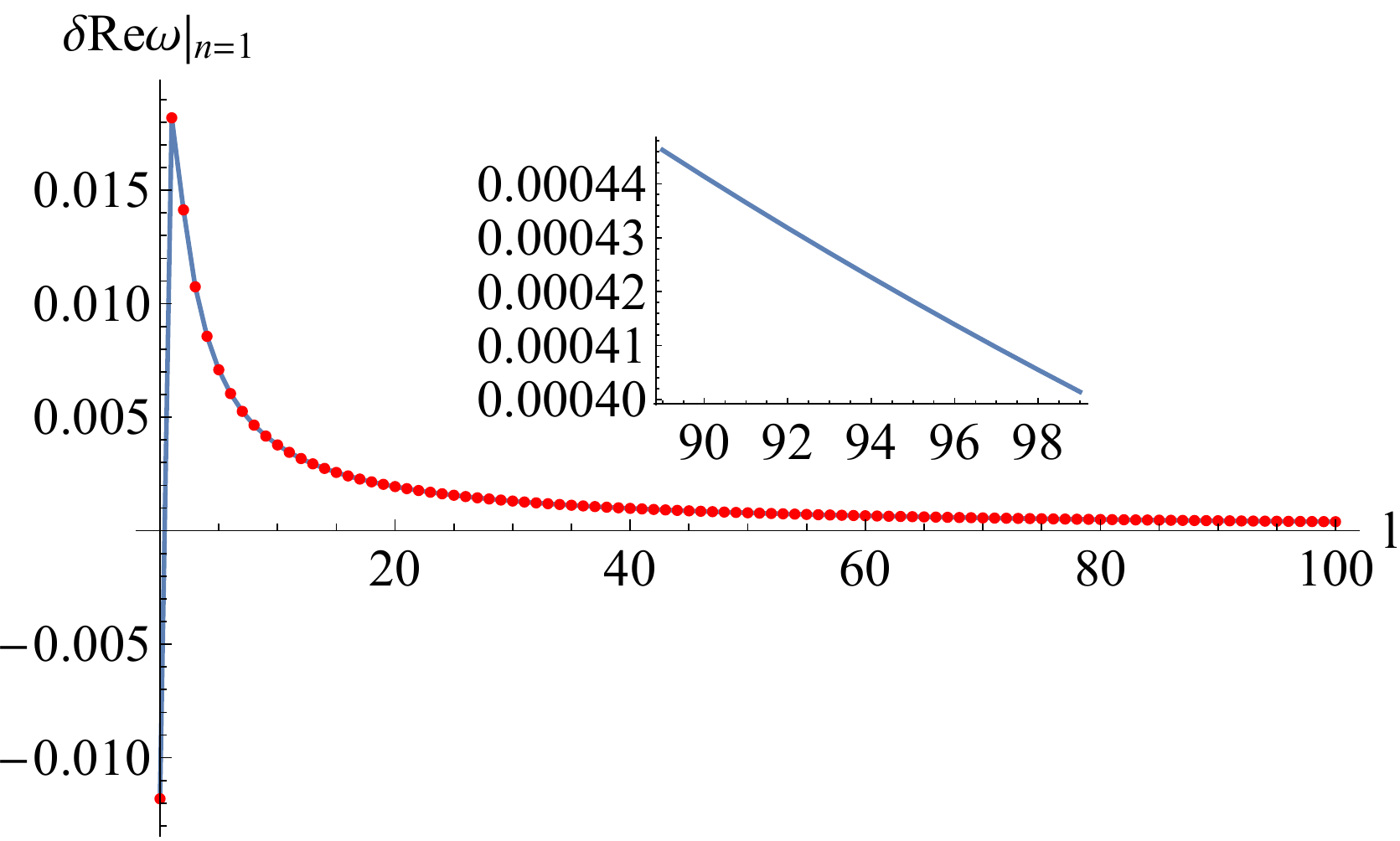}
		\caption{The difference of QNFs of scalar field with $r_0=1/2$ between the LQG-BH and SS-BH. Left plot is for $n=0$ and right plot for $n=1$.}
		\label{eikonal_scalar}
	}
\end{figure}

We notice that since the real part of QNF is independent of the LQG parameter $r_0$ in the eikonal limit. Therefore, we expect that as $l$ increases, the difference in $Re\omega$ between LQG-BH and SS-BH will be suppressed and vanish. Fig.\ref{eikonal_scalar} validates this argument that as $l$ increases, the difference rapidly decreases and goes to zero.

\section{Ringdown waveform}\label{sec-ring}

In this section, we will study the time evolution of the scalar and electromagnetic perturbations, which help us to further know the total contributions from overtones. Here, we will use the finite difference method (FDM) technics to implement the dynamical evolution. For more details on the FDM, we can refer to Refs.\cite{Lin:2022owb,Zhu:2014sya,Abdalla:2010nq,Yang:2021yoe,Fu:2022cul} and references therein. To this end, we write the wave equation in difference form as
\begin{eqnarray}
	-\frac{(\Psi_{i+1,j}-2\Psi_{i,j}+\Psi_{i-1,j})}{\triangle t^2}+\frac{(\Psi_{i,j+1}-2\Psi_{i,j}+\Psi_{i,j-1})}{\triangle r_*^2}-V_j \Psi_{i,j}+\mathcal{O}(\triangle t^2)+ \mathcal{O}(\triangle r_*^2)=0 \,, \nonumber \\
\end{eqnarray}
where $\triangle t$ and $\triangle r_*$ are the time and radial intervals, respectively, wihch are defined by $t=i \triangle t$ and $r_*=j \triangle r_*$. The $V_j$ is the discrete form of the effective potential \eqref{V_eff}. Then, the iterate formula is derived as:
\begin{eqnarray}
	\Psi_{i+1,j}=-\Psi_{i-1,j}+\frac{\triangle t^2}{\triangle r_*^2}(\Psi_{i,j+1}+\Psi_{i,j-1})+(2-2\frac{\triangle t^2}{\triangle r_*^2}-\triangle t^2 V_j)\Psi_{i,j}\,.
	\label{it-process}
\end{eqnarray}
Notice that the Courant-Friedrichs-Lewy (CFL) condition for instability requires that $\triangle t/\triangle r_*<1$.
Using the iterate formula \eqref{it-process} with the initial Gaussian distribution $\Psi(r_*,t<0)=0$ and $\Psi(r_*,t=0)=\exp{-\frac{(r_*-a)^2}{2b^2}}$, one can obtain the ringdown profiles.

In general, there are three different stages in time-evolution profile: initial outburst, quasinormal ringing, which depends only on the black hole's characteristics and is very important for GW observations \cite{LIGOScientific:2016aoc,Konoplya:2011qq,Kokkotas:1999bd,Berti:2009kk}, and the late tail, which exhibits the power-law behavior for the asymptotically flat spacetimes or exponential behavior for asymptotically de-Sitter spacetimes. We will focus on the properties of the latter two stages in this section.

\begin{figure}[H]
	\center{
		\includegraphics[scale=0.58]{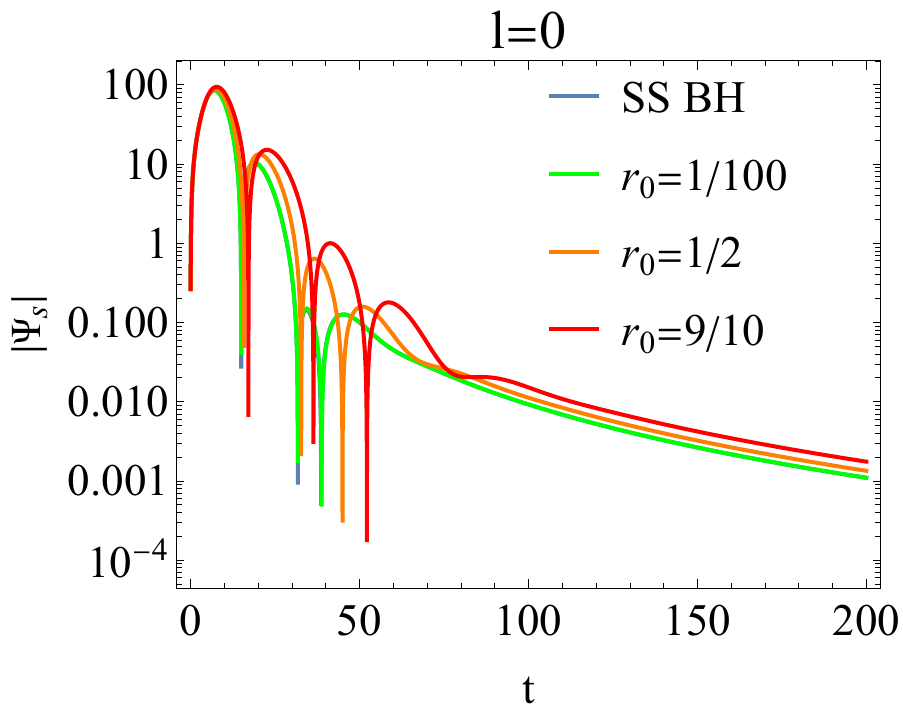}\hspace{0.05cm}
		\includegraphics[scale=0.58]{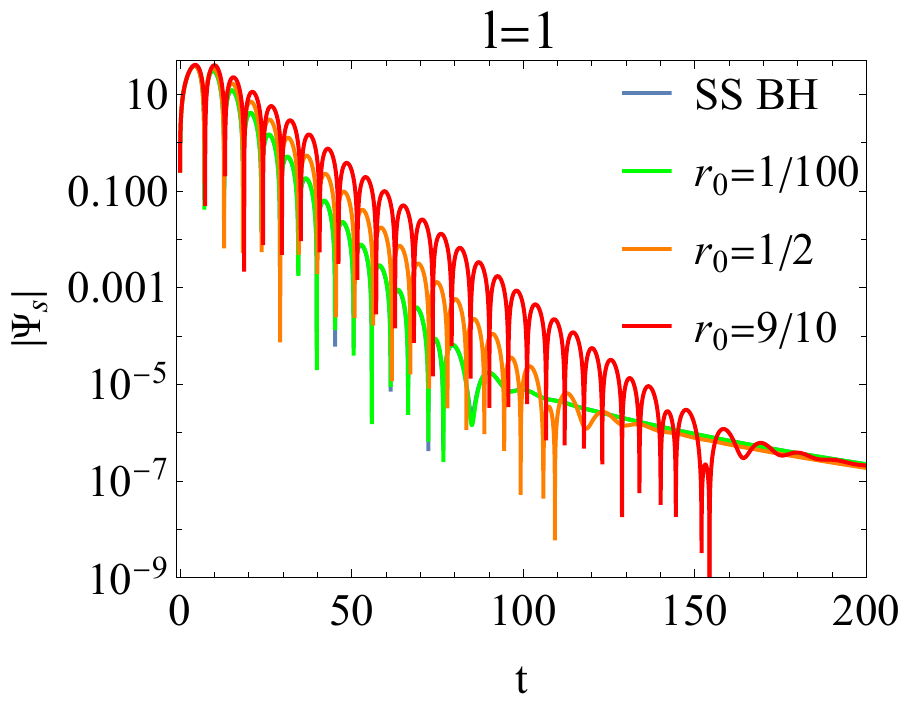}\hspace{0.05cm}		
		\includegraphics[scale=0.58]{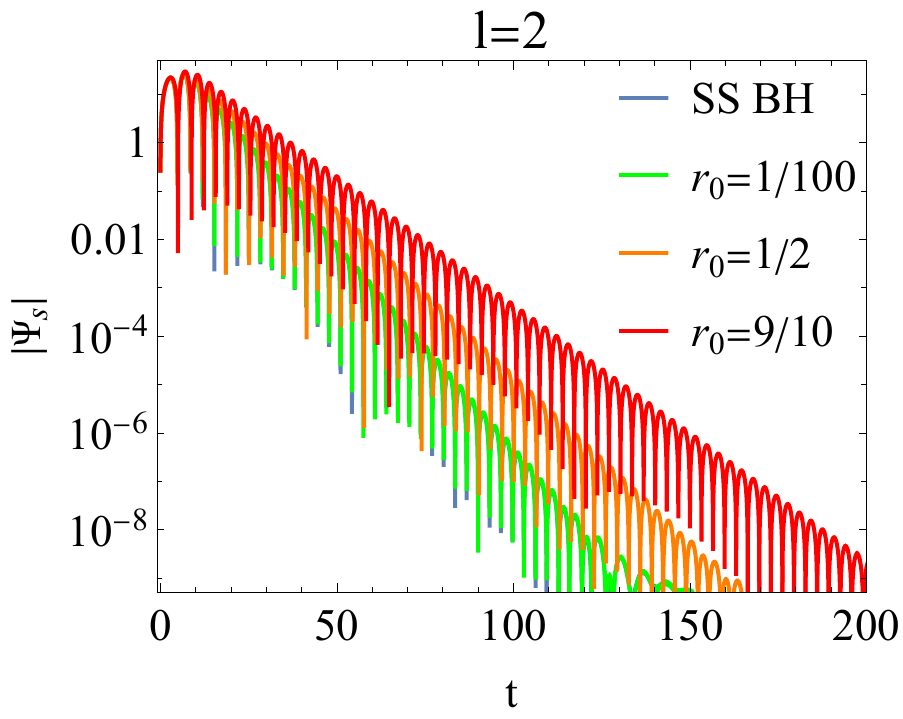}
		\caption{The semi-log plot of the time evolution of the scalar field $|\Psi_s(r)|$ for different $r_0$ with fixed $l$. Here, we have fixed the tortoise coordinate $r_*=5$.}
		\label{TD_scalar}
	}
\end{figure}

\begin{figure}[H]
	\center{
		\includegraphics[scale=0.8]{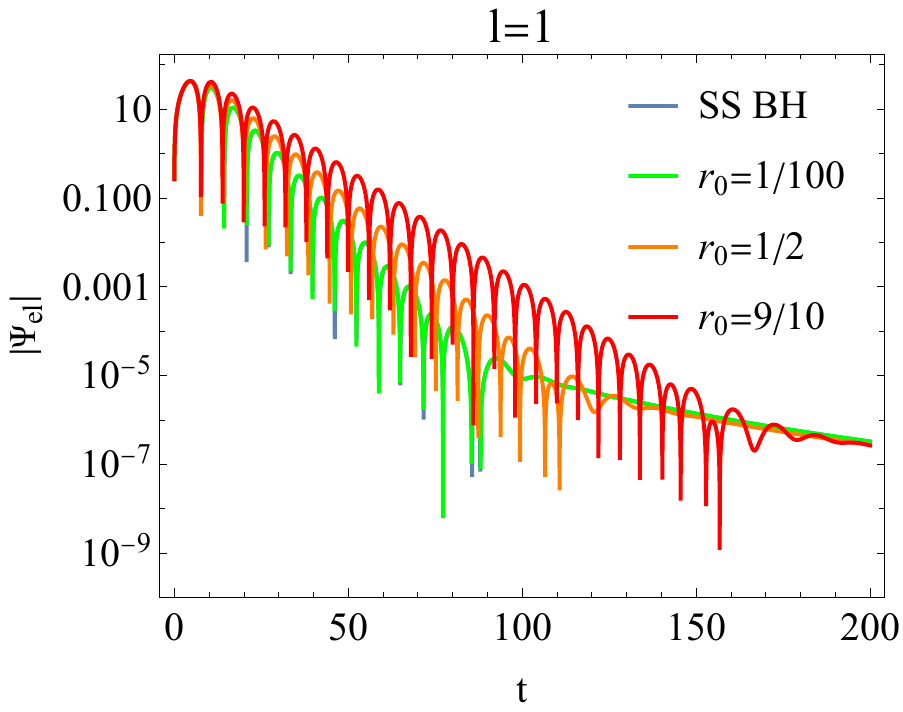}\hspace{0.1cm}
		\includegraphics[scale=0.8]{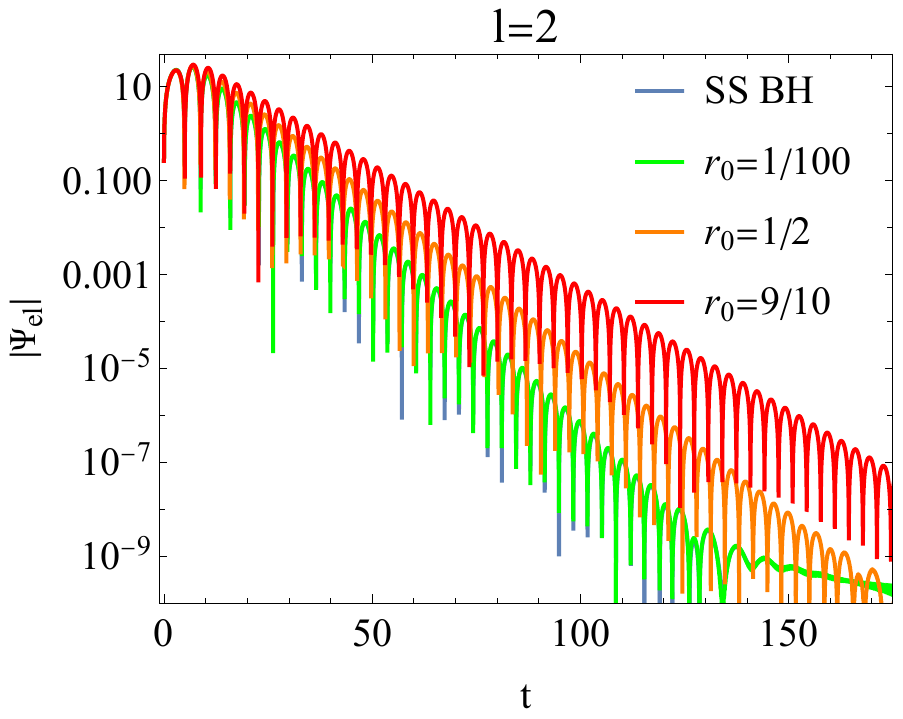}
		\caption{The semi-log plot of the time evolution of the electromagnetic field $|\Psi_{el}(r)|$ for different $r_0$ with fixed $l$. Here, we have fixed the tortoise coordinate $r_*=5$.}
		\label{TD_electro}
	}
\end{figure}

Figs.\ref{TD_scalar} and \ref{TD_electro} display the time-domain profile for both the scalar field and the electromagnetic field, respectively. In comparison to the SS-BH, the LQG-BH exhibit weaker oscillations and a slower decay rate during the intermediate time, primarily dominated by the fundamental mode. As expected, this observation aligns with the characteristics revealed by the fundamental mode presented in Tables \ref{TABLE-1}, \ref{TABLE-2}, \ref{TABLE-3} and \ref{TABLE-4}. In the asymptotically late-times, the quasinormal ringing is suppressed, and both LQG-BH and SS-BH follow the same power-law tail as $\Psi(t)\sim t^{-(2l+3)}$ \cite{Gundlach:1993tp,Price:1972pw,Price:1971fb}.

In addition, we use the Prony method as described in \cite{Zhidenko:2009zx} to calculate the fundamental mode. We will select the data from the time-domain profile during the intermediate time. We have presented the results for both the scalar field and electromagnetic field in Tables \ref{TABLE-5} and \ref{TABLE-6}, respectively. These findings align with the outcomes obtained by directly solving the eigenvalue problem, as presented in Talbes \ref{TABLE-1}, \ref{TABLE-2} \ref{TABLE-3} and \ref{TABLE-4}. In theory, we can fit the higher overtone QNFs after subtracting the contribution of the fundamental mode if the quasinormal ringing stage is sufficiently prolonged. Nevertheless, in real-world applications, fitting the higher overtones can be challenging due to their rapid damping, causing them to become nearly indistinguishable from numerical errors. We intend to undertake this analysis in the near future.

\begin{table*}[htp]
	\centering
	\begin{tabular}{c|cccc}
		\hline l & $\omega$  ($r_0=0$)  & $\omega$  ($r_0=1/100$)& $\omega$  ($r_0=1/2$)  & $\omega$  ($r_0=9/10$)   \\
		\hline
		0 & 0.22114-0.21068i & 0.22055-0.20984i & 0.20155-0.16416i & 0.17251-0.13331i 
		\\
		\hline
		1 & 0.58588-0.19513i & 0.58577-0.19443i & 0.57966-0.15821i & 0.57122-0.12296i 
		\\
		\hline
	\end{tabular}
	\caption{The fundamental modes of the scalar field perturbation  for various values of $l$ and $r_0$, which are determined using the prony method.\label{TABLE-5} }
\end{table*}
\begin{table*}[htp]
	\centering
	\begin{tabular}{c|cccc}
		\hline l & $\omega$  ($r_0=0$)  & $\omega$  ($r_0=1/100$)& $\omega$  ($r_0=1/2$)  & $\omega$  ($r_0=9/10$)   \\
		\hline
		1 & 0.49626-0.18372i & 0.49660-0.18319i & 0.51305-0.15239i & 0.52317-0.11786i 
		\\
		\hline
		2 & 0.91495-0.19021i & 0.91516-0.18959i & 0.92474-0.15566i & 0.93038-0.12032i
		\\
		\hline
	\end{tabular}
	\caption{The fundamental modes of the electromagnetic field perturbation  for various values of $l$ and $r_0$, which are determined using the prony method. \label{TABLE-6}}
\end{table*}

\section{Conclusion and discussion}\label{sec-conclusion}

As the rapid development of the GW detection technics, it is expected to detect the quantum gravity effect. To extract substantial information from GW detectors, one must thoroughly know the main features and behaviors of QNM for LQG-BH. As the first step, we investigate the QNM for both the scalar field and the electromagnetic field over the covariant LQG-BH proposed in \cite{Alonso-Bardaji:2021yls, Alonso-Bardaji:2022ear}.

First, we focus on the fundamental modes. It is found that the system is always stable under scalar field or electromagnetic field perturbations, and the LQG effect results in faster decaying modes. The difference is that the LQG effect reduces the oscillations in the scalar field, however it enhances oscillations in the electromagnetic field. In addition, we find that the system under the scalar field perturbation enjoys faster decaying modes with more oscillations than the electromagnetic field.

Some peculiar phenomena emerge in the QNM spectra with higher overtones. A noteworthy characteristic is the pronounced deviation observed in the first several overtones from their Schwarzschild counterparts, with this deviation increasing as the overtone number rises. Particularly noteworthy is the case of the scalar field with $l=0$, where the real oscillation frequency of the second overtone decreases by more than six times its Schwarzschild limit, while the fundamental mode undergoes only minor changes. This observation is also reflected in the trajectory within the $\omega_R$-$|\omega_I|$ phase diagram, which spirals towards a stable point. This emergence of the outburst of overtones can be attributed to disparities in the region near the event horizon between the Schwarzschild BH and the LQG corrected BH. Another notable phenomenon in higher overtones is the presence of oscillatory behavior in the QNFs of the scalar field with $l=0$ as the quantum parameter $r_0$ experiences a significant increase. These oscillations become more pronounced with a higher overtone number. This oscillatory pattern may be associated with the extremal effect.

Finally, we comment some open questions deserving further exploration.
\begin{itemize}
	\item It would be interesting to extend our investigation to the Dirac field and see if the peculiar property still emerges in the QNM spectra.
	\item It is definitely interesting and valuable to further study the QNM spectrum of the gravity perturbations. It provides us a platform for detecting quantum gravity effects using the GW detector. In addition, we can examinate if the isospectrality still holds in this LQG-BH model.
	\item In \cite{Lagos:2020oek}, the anomalous decay rate of QNMs of a massive scalar field is observed. Depending on how large the mass of the scalar field is, the decay timescales of the QNMs either grow or decay with increasing angular harmonic numbers. This anomalous behavior is seen in much larger class models beyond a simple massive scalar field, see \cite{Aragon:2020tvq,Aragon:2020teq,Fontana:2020syy,Gonzalez:2022upu,Gonzalez:2022ote} and references therein. It will interesting to see how the LQG effect affects this anomalous behavior.
	\item We can also construct an effective rotating LQG-BH solution using the Newman-Janis algorithm, starting with this spherical sysmetric LQG-BH, and study the LQG effects on its QNM spectrum and shadow, allowing us to constrain the LQG parameters using the GW detector and the Event Horizon Telescope (EHT).
\end{itemize}
We plan to investigate these questions and publish our results in the near future.

\acknowledgments

This work is supported by National Key R$\&$D Program of China (No. 2020YFC2201400), the Natural Science
Foundation of China under Grants No. 12375054 and No. 12375055, the Postgraduate Research \& Practice Innovation Program of Jiangsu Province under Grant No. KYCX20\_2973, the Postgraduate Scientific Research Innovation Project of Hunan Province under Grant No. CX20220509, the Science and Technology Planning Project of Guangzhou (202201010655), the Fok Ying Tung Education Foundation under Grant No. 171006, the Natural Science Foundation of Jiangsu Province under Grant No.BK20211601. J.-P.W. is also supported by Top Talent Support Program from Yangzhou University.

\appendix

\section{Wave equations}\label{app-wave-eq}

In this appendix, we will derive the wave equations for the scalar and electromagnetic fields in detail. First, we shall provide a generic version of the wave equation in a static spherically symmetric spacetime. The cases of scalar field and electromagnetic field are then discussed in detail.

Because the spacetime is static spherically symmetric, we can separate variables using the spherical function and write the radial equation in the form
\begin{eqnarray}\label{SL_form}
(\mathcal{K}(r) \mathcal{S}(r) \hat{\Psi}'(r))'+\Big(\Lambda \mathcal{F}(r)+ \mathcal{K}(r)\frac{\omega^2}{\mathcal{S}(r)}\Big)\hat{\Psi}(r)=0\,,
\end{eqnarray}
where $\hat{\Psi}$ is the radial part of the wave function, the coefficient functions $\{\mathcal{K}\,, \mathcal{F}\,, \mathcal{S}\}$ only depend on the radial coordinate $r$, and $\Lambda$ is the separation constant.
After introducing the tortoise coordinate $r_*$ and redefining the wave function as
\begin{eqnarray}\label{Sch_trans}
	\frac{dr_*}{dr}=\frac{1}{\mathcal{S}(r)}\,,\  \ \ \ \ \  \hat{\Psi}(r)=\frac{\Psi}{\sqrt{\mathcal{K}(r)}}\,,
\end{eqnarray}
Eq.\eqref{SL_form} can be recasted into the following form
\begin{eqnarray}
\frac{d^2\Psi(r_*)}{dr_*^2}+(\omega^2-V_{eff}(r_*))\Psi (r_*)=0\,.
\end{eqnarray}
The above formula provides a general transformation from the usual wave equation to its Schr\"{o}dinger-like counterpart.

In the following, we will go over the specific form of the wave equations for scalar and electromagnetic fields.
For the scalar field equation, we perform the separation as $\Phi(t,r,\theta , \phi)=\sum_{l,m}\hat{\Psi}(r)e^{-i \omega t}Y_{lm}(\theta ,\phi)$, where $Y_{lm}(\theta ,\phi)$ is the spherical harmonics.
When the particular form of the LQG-BH background \eqref{metric} is substituted into the wave equation \eqref{SL_form}, one obtains
\begin{eqnarray}
	\label{SL_form_scalar}
\left(r^2 f(r) \sqrt{g(r)} \hat{\Psi}'(r)\right)'+\left(\frac{r^2\omega^2}{f(r)\sqrt{g(r)}}-\frac{l(l+1)}{\sqrt{g(r)}}\right)\hat{\Psi}(r)=0\,.
\end{eqnarray}
We can read off the coefficient functions by comparing Eq.\eqref{SL_form_scalar} to Eq.\eqref{SL_form}
\begin{eqnarray}
	\label{coe-fun-scalar}
	\mathcal{K}(r)=r^2 \,, \ \ \mathcal{S}=f(r)\sqrt{g(r)}\,.
\end{eqnarray}
The Schr\"{o}dinger-like version of the wave equation is then easily given as
\begin{eqnarray}
&&
\label{s-scalar-app}
\frac{\partial ^2 \Psi}{\partial r_{*}^2}+(\omega^2 -V_s) \Psi=0\,,
\
\\
&&
\label{V-scalar-app}
 V_s=f(r)\frac{l(l+1)}{r^2}+\frac{1}{2r}\frac{d}{dr}f(r)^2g(r)\,.
\end{eqnarray}

For the electromagnetic field, we can expand the gauge field $A_\mu$ in vector spherical harmonics \cite{Ruffini,Cardoso:2003pj},
\begin{eqnarray}
	A_{\mu}(t, r, \theta, \phi) & = & \sum_{l, m}\left(\left[\begin{array}{c}
		0 \\
		0 \\
		\frac{a_{l m}(r)}{\sin \theta} \partial_{\phi} Y_{l m} \\
		-a_{l m}(r) \sin \theta \partial_{\theta} Y_{l m}
	\end{array}\right]+\left[\begin{array}{c}
		p_{l m}(r) Y_{l m} \\
		h_{l m}(r) Y_{l m} \\
		k_{l m}(r) \partial_{\theta} Y_{l m} \\
		k_{l m}(r) \partial_{\phi} Y_{l m}
	\end{array}\right]\right)e^{-i \omega t}\,,
\end{eqnarray}
where the first term is the odd (axial) perturbation and second term is even (polar) perturbation. Then, in the following, we will show how to derive the odd perturbation equation and even perturbation equation.

When we switch on the odd electromagnetic field perturbation, we can explicitly write down the radial equation as
\begin{eqnarray}
	\left(f(r) \sqrt{g(r)} a_{l m}'(r)\right)'+\left(\frac{\omega^2}{f(r)\sqrt{g(r)}}-\frac{l(l+1)}{r^2\sqrt{g(r)}}\right)a_{l m}(r)=0\,,
\end{eqnarray}
It is easy to find that $\mathcal{K}=1$ and $\mathcal{S}=f(r)\sqrt{g(r)}$. Thus, we have
\begin{eqnarray}
	V_{odd}=f(r)\frac{l(l+1)}{r^2}\,,
\end{eqnarray}
where $\Psi=a_{l m}(r)$.

For the even perturbation of the electromagnetic field, the radial equation becomes
\begin{eqnarray}\label{even_eq}
	&&
	p_{l m}''(r)+ q(r) p_{l m}'(r)+i \omega \left(h_{l m}'(r)+q(r)h_{l m}(r)\right)+\frac{l(l+1)}{r^2 f(r)g(r)}(p_{lm}(r)+ i \omega k_{lm}(r))=0\,, \ \nonumber \\
	&&
	-i \omega p_{l m}'(r) +\omega^2 h_{l m}(r)+\frac{l(l+1)}{r^2}f(r)(-h_{l m}(r)+k_{l m}'(r))=0\,,
\end{eqnarray}
where $q(r)=\frac{2}{r}+\frac{g'(r)}{2 g(r)} $. After introducing a new variable
\begin{eqnarray}
\hat{\Psi}(r)=-p_{lm}'(r)-i\omega h_{lm}(r)\,,
\end{eqnarray}
Eq.\eqref{even_eq} can be reduced to
\begin{eqnarray}
(r^4f(r)g(r)^{3/2}\hat{\Psi}'(r))'+\left(\frac{r^4\omega^2\sqrt{g(r)}}{f(r)}-l(l+1)r^2\sqrt{g(r)+\frac{1}{2}J(r)}\right)\hat{\Psi}(r)=0
\end{eqnarray}
where $J(r)=r^2\sqrt{g(r)}(rf'(r)(4g(r)+rg'(r))+f(r)(4g(r)+r(6g'(r)+rg''(r))))$. Thus, the coefficient functions are $\mathcal{K}=r^4 \sqrt{g(r)}$ and $\mathcal{S}=f(r)\sqrt{g(r)}$ and then we have
\begin{eqnarray}
	V_{even}=f(r)\frac{l(l+1)}{r^2}\,.
\end{eqnarray}
We find that the effective potentials for odd and even electromagnetic field perturbations are the same. Therefore, we will use $V_{el}$ to signify the effective potential of the electromagnetic field rather than $V_{odd}$ and $V_{even}$.

\section{QNMs in the eikonal limit}\label{app-QNM-eikonal}

In this appendix, we will show the connection between the QNMs in the eikonal limit and the behavior of null particle trapped on the unstable circular geodesic.
For a null particle, the Lagrange is\footnote{For the calculation details of the geodesic of a null particle, please refer to \cite{Chandrasekhar,Cardoso:2008bp,Perlick:2015vta,Wei:2019jve}.}
\begin{eqnarray}\label{H_J_lag}
	\mathcal{L}(x, \dot{x})=1/2 g_{\mu\nu}\dot{x}^\mu \dot{x}^\nu\,.
\end{eqnarray}
We start with the spherically symmetric geometry \eqref{metric}.
Thanks to the symmetry, one can only consider the geodesics in the equatorial plane: $\theta=\pi/2$. Then the Lagrangian \eqref{H_J_lag} becomes
\begin{eqnarray}
	\label{L-v2}
	2\mathcal{L}= -f(r)\dot{t}^2+\frac{\dot{r}^2}{f(r)g(r)}+r^2\dot{\phi}^2\,,
\end{eqnarray}
where the dot represents the derivative with respect to the affine parameter $\tau$. In this system, there are two constants of the motion, which are
\begin{eqnarray}
	\label{pt-pphi}
	P_t=-f(r)\dot{t}=-E \,, \ \ \ \ P_\phi=r^2\dot{\phi}=L\,.
\end{eqnarray}
Using the canonical transform and combining the above equations \eqref{L-v2} and \eqref{pt-pphi}, we have the following reduced Hamiltonian system:
\begin{eqnarray}
	2\mathcal{H}=E \dot{t}+\frac{\dot{r}^2}{f(r)g(r)}+L \dot{\phi}\,.
\end{eqnarray}
Since the Hamiltonian $\mathcal{H}$ satisfies the constraint $\mathcal{H}=0$, we have
\begin{eqnarray}\label{H-constraint}
\dot{r}^2+V_{eff}=0\,,
\end{eqnarray}
where the effective potential is
\begin{eqnarray}\label{Veff_PS}
	V_{eff}=g(r)\left(\frac{L^2}{r^2}f(r)-E^2\right)\,,
\end{eqnarray}
Because $\dot{r}^2>0$, the photon can only emerge in the area of negative potential.
When the angular momentum is small, the photon will fall from infinity into the black hole. However, for the large angular momentum, the photon will escape the bondage of the black hole and go back to infinity. Therefore, the critical circular orbit for the photon can be derived from the unstable conditions
\begin{eqnarray}
	V_{eff}=0 \,, \ \ \ \frac{\partial V_{eff}}{\partial r}=0 \,, \ \ \ \frac{\partial^2 V_{eff}}{\partial r^2}<0\,.
\end{eqnarray}
From the above conditions, we can obtain the equation for the critical radius $r_c$
\begin{eqnarray}\label{PS_cond2}
	2f_c(r)=r_c f_c'(r)\,.
\end{eqnarray}
Correspondingly, we have the critical impact parameters $b_c$:
\begin{eqnarray}\label{PS_cond1}
	b_c=\frac{L}{E}=\frac{r_c}{\sqrt{f_c(r)}}\,.
\end{eqnarray}
Then, the shadow radius $R_s$ and Lyapunov exponents $\lambda$ can be calculated as follows:
\begin{eqnarray}
	&&
	R_s=\sqrt{\zeta^2+\eta^2}=b_c=3\sqrt{3}m \,,
	\
	\\
	&&
	\lambda=\sqrt{\frac{V_{eff}''}{2 \dot{t}^2}}=\sqrt{-\frac{r_c^2}{f(r_c)}\left(\frac{d^2}{dr_*^2}\frac{f(r)}{r^2}\right)\Big{|}_{r=rc}}\,,
\end{eqnarray}
where $\{\zeta \,, \ \eta\}$ are the celestial coordinates. We find that the shadow radius reduces to the one of SS-BH \cite{Blome,Churilova:2019jqx}. It means that the LQG effect doesn't change the shape of the shadow. However, the LQG correction affects the Lyapunov exponent $\lambda$.

On the other hand, we shall use the first order WKB approximation to obtain the analytic form of the QNMs in the eikonal limit ($l \to \infty$). In this limit, the last term of the effective potential \eqref{V_eff} can be ignored, resulting in the following form of the effective potential
\begin{eqnarray}\label{Veff_limit}
	V_{\infty}(r)=f(r)\frac{l^2}{r^2}\,.
\end{eqnarray}
Reminding that the potential \eqref{Veff_PS} and \eqref{Veff_limit} are the same. Therefore, in the eikonal limit, the QNMs may be obtained by the multiples of the frequency and the instability timescale of the unstable circular null geodesic \cite{Cardoso:2008bp}:
\begin{eqnarray}\label{QNMs_eikonal-app}
	\omega=\Omega_c l-i(n+\frac{1}{2})|\lambda|\,,
\end{eqnarray}
where $\Omega_c$ is the angular velocity and can be worked out as
\begin{eqnarray}
	\Omega_c=\frac{\dot{\phi}}{\dot{t}}=\frac{1}{b_c}\,.
\end{eqnarray}

\bibliographystyle{style1}
\bibliography{Ref}
\end{document}